\documentclass[journal,12pt,draftclsnofoot,onecolumn]{IEEEtran}

\usepackage{comment}
\usepackage{amssymb}
\usepackage{subfigure}
\usepackage{amsmath}
\usepackage{amsfonts}
\usepackage{mathrsfs}
\usepackage{engord}
\usepackage[dvips]{graphicx}
\usepackage{bbm}
\usepackage{threeparttable}
\usepackage{booktabs}
\usepackage{epsfig}
\usepackage{color}
\usepackage{graphicx}
\usepackage{multirow}
\usepackage[sort&compress]{natbib}
\bibpunct{[}{]}{,}{n}{,}{,}
\begin{document}

\newtheorem{thm}{Theorem}
\newtheorem{prop}{Proposition}
\newtheorem{lem}{Lemma}
\newtheorem{method}{Coding method}
\newtheorem{remark}{Remark}[section]

\title{Partition Information and its Transmission over Boolean Multi-Access Channels}

\author{Shuhang Wu, Shuangqing Wei, Yue Wang, Ramachandran Vaidyanathan and Jian Yuan}
\maketitle

\footnotetext[1]{S. Wu, Y. Wang and J. Yuan is with Department of Electronic Engineering, Tsinghua University, Beijing, P. R. China, 100084. (E-mail: {wsh05}@mails.tsinghua.edu.cn; {wangyue, jyuan}@mail.tsinghua.edu.cn). S. Wei and R. Vaidyanathan are with the School of Electrical Engineering and Computer Science, Louisiana State University, Baton Rouge, LA 70803, USA (Email: swei@lsu.edu, vaidy@lsu.edu). }

\begin{abstract} 
	In this paper, we propose a novel partition reservation system to study the partition information and its transmission over a noise-free Boolean multi-access channel. The objective of transmission is not message restoration, but to 
partition active users into distinct groups so that they can,
subsequently, transmit their messages without collision.  We first calculate
(by mutual information)
the amount of information needed for the partitioning without channel effects,
and then propose two different coding schemes to obtain achievable transmission rates over the channel.
The first one is the brute force method, where the
codebook design is based on centralized source coding; the second method uses random coding where the codebook is generated randomly and optimal Bayesian decoding is employed to reconstruct the partition. Both methods shed light on the internal structure of the partition problem. A novel hypergraph formulation is proposed for the random coding scheme, which intuitively describes the information in terms of a strong coloring of a hypergraph induced by a sequence of channel operations and interactions between active users. An extended Fibonacci structure is found for a simple, but non-trivial, case with two active users. A comparison between these methods and group testing is conducted to demonstrate the uniqueness of our problem.
\end{abstract}

\begin{IEEEkeywords}
partitioning information, conflict resolution, Boolean algebra, Fibonacci numbers.
\end{IEEEkeywords}

\section{Introduction}
	One primary objective of many coordination processes is to order
a set of participants. For example, multiaccess can be viewed as 
(explicitly or implicitly) ordering a set of users for exclusive access to 
a resource.  Information interaction plays a key role
in establishing such an order.
To formalize this interactive information and derive fundamental limits on its
transmission, we propose in this paper a novel partition reservation model over
a noise-free Boolean multi-access channel and use an information theoretic approach in
its analysis.

For the simplest variant of the problem we study,
let $\mathcal{N}=\{1,\ldots,N\}$ be a set of $N$ users and let $\mathcal{G}_{\mathbf{s}}=\{i_1,\ldots, i_K\}\subseteq\mathcal{N}$ be a set of $K$ {\it active} users. The problem is to let all users obtain a common ordered $K$-\emph{partition}\footnote{An ordered $K$-partition $\Pi = (\mathcal{B}_1, \ldots, \mathcal{B}_K)$ of $\mathcal{N}$ is a sequence of $K$ non-empty subsets of $\mathcal{N}$ that satisfies the following conditions: (a) for all $1\leq i < j \leq K$, $\mathcal{B}_i \cap \mathcal{B}_j = \emptyset$ and (b) $\bigcup_{i=1}^K \mathcal{B}_i = \mathcal{N}$.} $\Pi = (\mathcal{B}_1, \ldots, \mathcal{B}_K)$ of $\mathcal{N}$, 
so that each group (or block) $\mathcal{B}_i$ has exactly one active user from $\mathcal{G}_{\mathbf{s}}$. Equivalently, we use a vector $\mathbf{z} = [z_1, \ldots, z_K]^\top$ to represent the ordered $K$-partition $\Pi$, where $z_i\in \mathcal{K} \triangleq \{1,2,\cdots,K\}$ is the id of the group that user $i$ belongs to, i.e., $i \in \mathcal{B}_k$ iff $z_i = k$. The desired partition is determined by a series of transmissions and observations over a channel, more specially, users are connected through a shared slotted Boolean multi-access channel. Suppose that during slot~$t$, each active user~$i$ transmits bit $x_{i,t}\in\{0,1\}$ on the channel. A common feedback $\displaystyle{y_{t}=\bigvee_{i\in \mathcal{G}_{\mathbf{s}}} x_{i,t}}$ will be observed by all users, i.e., if no active users transmit bit 1 during slot $t$, $y_t=1$; if at least one active user transmits 1, $y_t = 0$. 
The goal is to previously schedule $T$ rounds of transmissions (denoted by an accessing matrix 
$\displaystyle{\mathbf{X}\triangleq [x_{i,t}]_{1\leq i \leq N, 1\leq t \leq T}}$),
and a common decoding function $g(\cdot)$, so that after observing $T$ rounds feedbacks $\mathbf{y}\triangleq [y_1,\ldots, y_T]^{\top}$, a desired ordered $K$-\emph{partition} of $\mathcal{N}$, denoted by $\mathbf{z} = g(\mathbf{y})$ can be obtained by all users.
The objective is to find an
achievable lowerbound on the number of slots $T$, within which there exists a matrix $\mathbf{X}$ and $g(\cdot)$ so that every possible active set $\mathcal{G}_{\mathbf{s}}\subseteq \mathcal{N}$ can be partitioned.



In the problem we consider, we do not seek to restore the states of all users
(that is, determine $\mathcal{G}_{\mathbf{s}}$ exactly), but to partition $\mathcal{G}_{\mathbf{s}}$ and to make users know the partition $\mathbf{z}$. Thus, a particular partition information that only pertains to the relationship between active users in $\mathcal{G}_{\mathbf{s}}$, is transmitted through the Boolean multi-access channel. We will formalize this
partition information, and focus on the achievable bound of its transmission rate over Boolean multi-access channels. This problem plays a significant role in understanding the fundamental limits on the capability of establishment of order in distributed systems.

	The proposed problem has a close relationship to a typical slotted conflict resolution problem \cite{1057022},  where each active users must transmit without conflict at least once during $T$ slots, i.e., if $x_{i,t}=1$ denotes
a trial of transmission for active user $i$ at slot~$t$, then there exists a $1\leq t_i \leq T$ such that $x_{i,t_i} = 1$, and for all $j\in\mathcal{G}_{\mathbf{s}}-\{i\}$, we have $x_{j,t_i} = 0$.
To achieve this goal, mainly two types of systems are studied: direct transmission system and reservation system with group testing \cite{1096146}. Direct transmission focuses on directly designing an $N \times T_{dr}$
accessing matrix ${\bf X}_{dt}$ (subscript $dt$ is for direct transmission), so that each node finds at least one slot for its exclusive access to the channel. Note that active users are implicitly partitioned during the transmission to ensure the success of transmission, (more specially, if the successful transmission time $t_i$ for each active user $i$ is known, the desired partition can be constructed by all the users), but the partition is not necessary to be known. Reservation with group testing has two stages. In the first reservation stage, an accessing matrix
$\mathbf{X}_{g}$ and decoding function $g(\cdot)$ are designed such that $\mathcal{G}_{\mathbf{s}}$ is exactly determined by $g(\mathbf{y})$, where $\mathbf{y}$ is the channel feedback. That is, (active or inactive) states of all users are restored and,
subsequently, active users can transmit in a predetermined order without conflict in a second
stage. The reservation stage is also called group testing \cite{ding2000combinatorial} or compressed sensing \cite{malyutov2013search} in different fields. The two stages can be of different time scales. We can see the payload transmission is separated, but in the reservation stage, $\mathcal{G}_{\mathbf{s}}$ is known to all users, which is more than we need.

	Compared with group testing and direct transmission system, our partitioning reservation system provides a new way to individually analyze the process of partitioning, which is the essence of coordination in conflict resolution problems.
It can be used as a reservation stage instead of group testing in conflict resolution problems, and holds the possibility of requiring fewer resources,
since it seeks only to partition $\mathcal{N}$, rather than restore $\mathcal{G}_{\mathbf{s}}$.
(Notice that once $\mathcal{G}_{\mathbf{s}}$ is restored, obtaining a partition
is straightforward.)
Compared with direct transmission,
we observe that usually, the time scale for reservation can be much smaller
in partition/reservation than that in payload transmission, thus it may need less time for conflict resolution in practical use. 

    The proposed partition reservation system has abundant applications in different areas. First, and foremost, it can be directly applied to the reservation stage in conflict resolution instead of group testing. Second, since the obtained common partition is known to all users in the partition reservation system, more complicated coordination among active users can be anticipated other than avoiding conflict in time domain, which is not attainable in traditional conflict resolution schemes. For example, code-division multiple accessing codes could be assigned to users in different groups based on the obtained partition, so that active users can claim accessing code sequences from a common pool in a distributed way without coordination from a central scheduler. Other examples can be found in parallel and distributed computation \cite{lynch1996distributed, xavier1998introduction, attiya2004distributed}, such as leader election \cite{hromkovic2005dissemination}, broadcasting \cite{Kowalski:2005:SPR:1073814.1073843}. In this paper, the system is constrained to a case with $K$ active users non-adaptively accessing a noiseless Boolean binary
 feedback channel. It is a fundamental case of the problem, but also has practical values. Consider a system with $N$ users each of which stays active with a probability $p$. If $p$ scales with $N$ such that the number of active users $K \approx N p$ is approximated as a constant for large $N$, the accessing of these $K$ active users to a Boolean channel is the case we are tackling. In our proposed framework, no adaptation is allowed over the Boolean channel, thereby reducing the expenditures on feedback overhead as compared with the adaptive models. It should be noted such non-adaptive channel model has also been considered in MAC or group testing literature (for example, \cite{capetanakis1979generalized, DeBonis2003223, sandor2008, ding2000combinatorial}, etc.). Our study will help us understand the fundamental limit on transmission resources to attain a partitioned coordination among the active users.

To achieve this goal, we first use source coding to quantify the partition
information. Then two coding schemes for the accessing matrix
$\mathbf{X}$, and decoding function $g(\cdot)$ are proposed. The  first
is a brute force method to design $\mathbf{X}$ and $g(\cdot)$ directly
based on results from source
coding. The purpose of source coding is to compress the source information, more
specifically, to find a set $\mathcal{C}$ of minimum number of partitions, so that for nearly any possible active 
set $\mathbf{s}$, there is a valid partition in $\mathcal{C}$. Then the brute force method tries to find the valid partition by checking every partition in $\mathcal{C}$ using the channel.
 The second scheme, employing random coding, generates
accessing matrix elements $x_{i,t}$ i.i.d. a by Bernoulli distribution,
then the partition is recovered by optimal Bayesian decoding.
The two methods can both work, and provide different views of this problem. In particular,
in the brute force method, if $T_{BF} = \frac{K^{K+1}}{K!}f(N)$ and $f(N)$ is an arbitrary function satisfying $\displaystyle{\lim_{N \to \infty} f(N) = \infty}$, the average error probability $P^{(N)}_e \leq e^{-f(N)} \to 0$, as $N \to \infty$. While for a simple but non-trivial $K=2$ case, we prove in random coding, if for any $\xi>0$, $\frac{log N}{T} \leq \displaystyle{\max_{0\leq p \leq 1} C(p)} - \xi$, where $C(p) = -(1-(1-p)^2)\log \varphi(p) -(1-p)^2\log (1-p)$, $\varphi(p) = \frac{p+\sqrt{4p-3p^2}}{2}$, we have the average error probability $P_e^{(N)} \leq \frac{1}{N^{\Delta}} \to 0$ for some $\Delta>0$, i.e., with polynomial speed. The two achievable bounds are shown better than that of group testing. 

	Moreover, for the random coding approach, we use a framework to solve
the problem from the view of strong coloring of hypergraphs, namely, the
partition objective can be transformed to the strong coloring problem
of a resulting hypergraph, and the effect of channel(s) is reflected
by a series of operations on hypergraph edges. Under such a framework,
the partition information is represented by types of hypergraphs in
which hyper-edges are determined by the interaction among a set of
possible active nodes. The joint work between the encoder and decoder
is to make sure that the resulting hypergraphs become strong colorable
after transmissions by active nodes and intervention by channels. In a
simple, but nontrivial, case with $K=2$ active users for a set of $N$ users, a suboptimal odd cycle based
analysis is proposed, and a structure of extended Fibonacci numbers is
found, which sheds lights on the inherent structure of the partition
information and Boolean channel, and could be further explored for $K>2$ cases. 

	As a summary, the contributions of this paper are twofold. First, we formulate
	a novel partition reservation problem which captures the
	transmission and restoration of some relationship information
	among active users. This relationship communication problem is
	also represented in a hypergraph based framework. Secondly,
	we propose two types of coding approaches, and the corresponding
	achievable bounds on the communication period, which provides
	the intuitive examples to study the relationship information
	transmission over Boolean multi-access channels.

	Part of our results has been presented in \cite{wsh2}. In this paper, we provide a more complete and comprehensive solution the formulated problems. In particular, we discuss the source coding problem in Section \ref{sec.sourcecoding}. Then based on source coding, we give a brute force coding method in Section \ref{sec5} to solve the partition problem. In Section \ref{sec7}, a sub-optimal decoding approach for the case of $K=2$ is provided which requires the resulting graph without odd cycles (i.e. two-colorable). Detailed proofs are then given to both Lemma 2 and Theorem~\ref{prop2}, which are not included in \cite{wsh2} due to space limitation. 

	The rest of this paper is organized as follows: in Section \ref{relatedworks}, we introduce the related work. The problem formulation appears in Section \ref{sec.formulation}. In Section \ref{sec.sourcecoding}, the partition information is illustrated by centralized source coding, then a brute force method directly inspired by source coding is proposed in Section \ref{sec5}. In Section \ref{sec6}, a random coding method is considered and the problem is reformulated in terms of a hypergraph. Based on this, a simple, but non-trivial, result for $K=2$ in random coding is analyzed in Section \ref{sec7}. In Section \ref{sec8}, we compare our results with that of group testing. We summarize our results and make some concluding remarks in Section \ref{sec9}.

\section{Related Work} \label{relatedworks}

	Although the proposed partition model could be useful in many problem settings, typical applications are in conflict resolution problems. The works on conflict resolution are too extensive to be included in our review here, and we thus only include those most relevant to our problem settings as described earlier. 

    To the best of our knowledge, Pippenger \cite{1056332} first expresses the nature of conflict resolution the mixture of two stages: (a) partitioning active users into different groups; (b) payload transmission. Hajak \cite{1056551} further studies this problem. In their model, $K$ users are randomly distributed (uniform or Poisson) in the $[0,1]$ real interval, denoted by $U=(U_1, \ldots, U_K)$, where $U_k$ is the arrival time of user $k$; a valid $K$-partition of $[0,1]$, denoted by $A$, should be done during the conflict resolution so that active users are separated in different groups. This model corresponds to our model when $N \to \infty$. By directly considering the mutual information between input $U$ and valid output $A$ (without consider the channel effect), a bound of throughput is derived to solve the conflict in an adaptive scheme over a $(0,\ldots, d)$-channel ($d\geq 2$), where the feedback is $0$ if no active users transmitted; 1, if only one user is active; ...; and $d$, if more than $d$ users are active. Minimum of $I(A; U)$ ($=\log \frac{K^K}{K!}$) was called the partition information by Hajak \cite{1056551}, which gave an achievable bound of a probabilistic problem. Suppose the elements of $U$ are uniformly distributed in $[0,1]$, and for a set of $K$-partitions $\{A_l\}_{l=1}^L$, let $P_{\{A_{l}\}}(K)$ be the probability of the event that at least one of the $A_l$ is a valid partition, and let $Q_{L}(K)$ be the minimum of $1-P_{\{A_{l}\}}(K)$ for different choice of $\{A_l\}_{l=1}^L$. Then, $Q_{L}(K) \leq (1-\frac{K!}{K^K})^L$. The lower bound of $Q_{L}(K)$ is discussed by Hajek, K\"{o}rner, Simonyi and Marton \cite{hajek1987conjectured, korner1988separating, 2639}, and seeking the tight lower bound still remains an open question\footnote{
K\"{o}rner gives $\frac{1}{L}\log \frac{1}{Q_{L}(K)} \leq \frac{K!}{K^{K-1}}$ by using graph entropy in \cite{korner1988separating}.}.

	This partition problem (without considering channel effect) is also closely related to perfect hashing, zero-error capacity, list codes, etc. \cite[Chap. V]{720537}. The problem is formulated in a combinatorial way: a subset $\mathcal{A}$ of $\mathcal{K}^L$ is called $K$-separated if every subset of $\mathcal{A}$ consisting of $K$ sequences is separated, i.e., if for at least one coordinate $i$, the $i$th coordinates of the said sequences all differ. Let $A_L = A_{L}(K)$ denote a maximal $K$-separated subset of $\mathcal{K}^L$. It can be seen that $A_L$ corresponds to $N$ users in our problem settings, and set $\mathcal{A}$ can be viewed as a set of $K$ partitions with size $L$, so that for any active set out of $A_L$ users, there exists a valid $K$ partition. The relationship between this combinatorial model and the probabilistic model is stated by K\"{o}rner \cite{korner1988separating}. Note that these problems did not consider the channel effect, thus, they were kind of source coding from information theoretic perspective. For completeness, we will state the source coding problem further in Section \ref{sec.sourcecoding} of this paper. In contrast, the problem we are focusing on is the transmission problem, i.e., construction of a valid partition relationship among active users by the feedback resulting from their explicit transmission over a collision Boolean multi-access channel. This problem has not been addressed previously, to the best of our knowledge.

In addition to the conflict resolution problems, there have been extensive works on  direct transmission and group testing that consider channel effects from the \emph{combinatorics} and \emph{probabilistic} perspectives. Ding-Zhu and Hwang provide in \cite{ding2000combinatorial} an overview; more specific approaches can be found
on superimposed codes for either disjunct or separable purposes
\cite{kautz1964nonrandom, dyachkov1983survey, DeBonis2003223, sebHo1985two, chen2007exploring},
on selective families \cite{Kowalski:2005:SPR:1073814.1073843},
on the broadcasting problem \cite{Clementi:2001:SFS:365411.365756}, and
for other methods \cite{capetanakis1979generalized, 1057020, sebHo1985two}. 
	It should be noted that recently, group testing has been reformulated using an information theoretic framework to study the limits of restoration of the IDs of all active nodes over Boolean multiple access channels \cite{6157065}.
We address in this paper the transmission of partition information (rather than identification information) over the channel, and it is thus, different from existing work.

\section{System model}\label{sec.formulation}

\subsection{Formulation}
	In this paper, lower-case (resp., upper-case) boldface letters are used for column vectors (resp., matrices).
For instance, $w_{i}$ is used for the $i$-th element of vector $\mathbf{w}$,
and $w_{i,t}$ is used for the $(i,j)$-th element of matrix $\mathbf{W}$.
Logarithms are always to base 2. The probability of a random variable $A$ having value $\tilde{A}$ is denoted by $p_A(\tilde{A}) \triangleq \text{Pr}(A=\tilde{A})$.  Similarly, $p_{A|B}(\tilde{A}|\tilde{B})\triangleq \text{Pr}(A=\tilde{A}|B=\tilde{B})$. Where there is no danger of ambiguity, we will drop the subscripts and simply write $p(A)$ or $p(A|B)$ to denote the above quantities.

	Assume the number of active users $K$ is known to  all users. The users are also given a common $N\times T$ accessing matrix (or codebook) $\mathbf{X}$, and a decoding function $g(\cdot)$. We use a Boolean vector $\mathbf{s}=[s_1, \ldots, s_N]^{\top}$ to represent the active or inactive states of users, where $s_i = 1$ iff
user~$i$ is active (that is, $i \in \mathcal{G}_{\mathbf{s}}$). Active users will use
$T$ slots to transmit according to codebook $\mathbf{X}$ and observe the feedback $\mathbf{y}=\left[y_t~:~1\le t\le T\right]^\top$ over these $T$ slots.
Then users derive the partition $\mathbf{z} = g(\mathbf{y})$. There are two dimensions in this problem, the user dimension of size $N$ and the time dimension
of size $T$.

{\it An Example}

\begin{figure}
\centering
\includegraphics[width=3.2in]{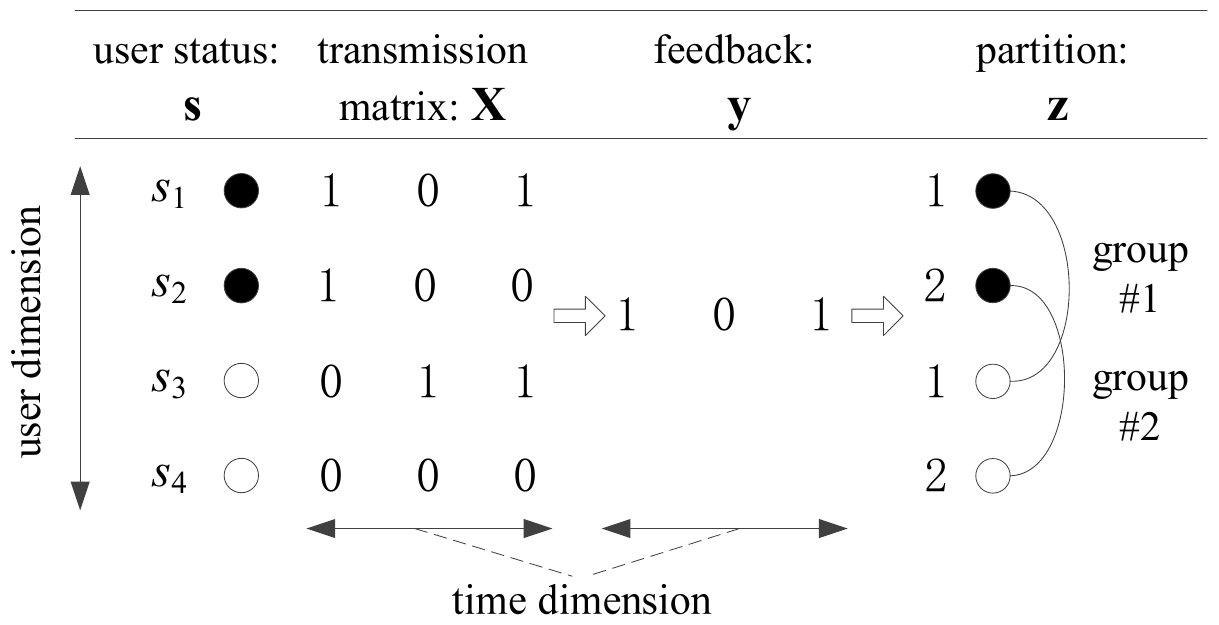}
\caption{Example of the formulation. ($N=4$, $K=2$, $\mathcal{G}=\{1,2\}$ indicates that users~1 and 2 are active; the total number of time slots is $T=3$. )} \label{pfig1}
\end{figure}

Our approach is illustrated by an example in Fig. \ref{pfig1} with four users
from $\mathcal{N}=\{1,2,3,4\}$, of which the users of set $\mathcal{G}_{\mathbf{s}}=\{1,2\}$ are active. The $N\times T$ codebook is
$\mathbf{X}$. 
In each slot $1\le t\le 3=T$, user~$i$ writes to the channel iff $i$ is active and $x_{i,t}=1$.
For example, in slot~1, that has $x_{1,1}=x_{2,1}=1$ and $x_{3,1}=x_{4,1}=0$,
both active users 1 and 2 write to the channel, resulting in a channel
feedback of $y_1=1$. In slot~2, $x_{3,2}=1$, however, since user~3 is
not active, there is no write and $y_2=0$. In slot~3, users 1 and 3 are called upon to write, but only user~1 writes as user~3 is not active.
The channel feedback over the three slots is $\mathbf{y}=\left[y_1,y_2,y_3\right]^\top=\left[1,0,1\right]^\top$. From this feedback, the knowledge of $K=2$
and the accessing matrix $\mathbf{X}$, the following conclusions can be drawn.
\begin{itemize}
\item Because $x_{3,2}=1$ and $y_2=0$, it can be concluded that user~3 is not active.
\item Because $x_{1,3}=x_{3,3}=1$ and $y_3=1$, it can be concluded that user~1
is active (as user~3 is inactive), also $\mathcal{G}_{\mathbf{s}} \nsubseteq \{2,4\}$.
\item The interaction in slot~1 only says that $\mathcal{G}_{\mathbf{s}} \nsubseteq \{3,4\}$.
\item Since $K$ is known to be 2, we conclude that exactly one of users 2 and 4 must be
active and the other inactive.
\item Thus partition $\{\{1,3\},\{2,4\}\}$ of $\mathcal{N}$ separates
active nodes into different groups, and $\mathbf{z}=[1~2~1~2]^\top$ can be selected as the result of decoding $\mathbf{y}$.
\end{itemize}
Observe that (unlike the restoration of $\mathcal{G}_{\mathbf{s}}$),
we do not (and need not) know which among users~2 and 4 is active.
Likewise although we happen to know that user~1 is active and user~3 is not,
this knowledge is coincidental; the partition approach does not invest
resources to seek this knowledge.


	To have a more general formulation, the problem can be treated as a coding problem in multi-access channels from the information theoretic view. Consider $N$ users whose active states are given in a vector $\mathbf{s} \in \mathbb{S}_{K;N} \triangleq \{\mathbf{s} \in \{0,1\}^N:\sum s_i = K\}$. The $i$-th row of $\mathbf{X}$, denoted by $\mathbf{x}_i^{\top}$ can be viewed as a codeword of user $i$ (note that $\mathbf{x}_i$ is a column vector, we would like to use a row vector $\mathbf{x}_i^{\top}$ to represent the codeword according to the tradition). It is also easy to see that
user~$i$ sends $s_i \mathbf{x}_i^{\top}$ on the channel. The channel feedback
is $\mathbf{y} = \bigvee_{i=1}^{N} s_i \mathbf{x}_i \triangleq [\bigvee_{i=1}^{N} s_i x_{i,t}]_{t=1}^T$. Then the decoded output is a partition\footnote{In the traditional view \cite{jordan2010mu}, a $K$ ordered partition of $\mathcal{N}$ is a $K$-tuple of subsets of $\mathcal{N}$, denoted by $(\mathcal{B}_1, \mathcal{B}_2, \ldots, \mathcal{B}_K)$, where $\forall 1\leq K_1 < K_2\leq K$, $\mathcal{B}_{K_1}\neq \emptyset$, $\mathcal{B}_{K_1} \cap \mathcal{B}_{K_2} = \emptyset$ and $\displaystyle{\bigcup_{k=1}^K \mathcal{B}_k = \mathcal{N}}$. Our notation here is equivalent. For example, for a partition denoted by $\mathbf{z} = [3~1~1~2~2]^\top \in \mathbb{Z}_{3;5}$, it represents a partition $(\{2,3\}, \{4,5\}, \{1\})$.} $\mathbf{z} \in \mathbb{Z}_{K;N}$, 
where:
\begin{align}
	\mathbb{Z}_{K;N} = \left\{\mathbf{z} \in \mathcal{K}^N:\forall 1\leq k \leq K, \exists z_i = k\right\} \nonumber
\end{align}
is the set of all possible $K$-ordered partition. A {\it distortion function} is defined for any
status vector $\mathbf{s}\in\mathbb{S}_{K;N}$ and a partition vector $\mathbf{z} \in \mathbb{Z}_{K;N}$ as follows:
\begin{align}
	d(\mathbf{s},\mathbf{z}) = 
	\begin{cases}
		0, &
			\mbox{if}~ \forall i,j \in \mathcal{G}_{\mathbf{s}},
			~~~~(i\neq j)\Longrightarrow (z_i\neq z_j)\\
		1, &\rm{otherwise}
	\end{cases}.
	\label{distortion1}
\end{align}
	The objective is to design a proper matrix $\mathbf{X}$ and a corresponding decoding function $\mathbf{z} = g(\mathbf{y})$, so that $d(\mathbf{s},g(\mathbf{y}))=0$ for nearly all $\mathbf{s} \in \mathbb{S}_{K;N}$. 


\begin{figure}
\centering
\includegraphics[width=3.2in]{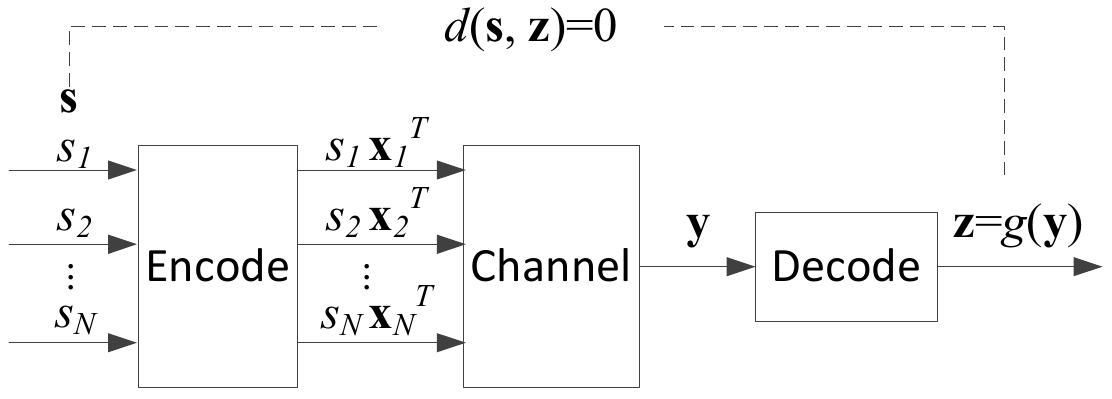}
\caption{Encoding-channel-decoding system with distortion criterion} \label{pfig2}
\end{figure}

	To simplify the notation, we write $\mathbf{y} = \mathbf{X}^\top\otimes \mathbf{s}$, where $\otimes$ denotes Boolean matrix multiplication in which
the traditional arithmetic multiplication and addition operations are 
replaced by logical AND and OR, respectively.
For any given $\mathbf{s}$, we denote the set of all possible desired
$\mathbf{z}$ as
$\mathbb{Z}_{K;N}(\mathbf{s}) = \left\{\mathbf{z} \in \mathbb{Z}_{K;N} : d(\mathbf{s},\mathbf{z}) = 0\right\}$.
	The set of all possible vectors $\mathbf{s}$ that are
compatible with a given $\mathbf{z}$ to produce 0 distortion is denoted by
	$\mathbb{S}_{K;N}(\mathbf{z}) = \{\mathbf{s} \in \mathbb{S}_{K;N} : d(\mathbf{s},\mathbf{z}) = 0\}$.
In some situations, we will need to know the number of users,
$n_k$, in a given group $k \in \mathcal{K}$. The set of all possible $\mathbf{z}$ with
group sizes $(n_1,\ldots, n_K)$, where 
$\displaystyle{\sum_{k=1}^K n_k =N}$, is denoted by:
\begin{align}
	\mathbb{Z}_{K;N}(n_1,\ldots, n_K) \triangleq \left\{\mathbf{z} \in \mathbb{Z}_{K;N}: \left(\sum_{i=1}^N \mathbbm{1}(z_i=k)\right) = n_k, 1\leq k \leq K\right\},
\nonumber
\end{align}
here the indicator function $\mathbbm{1}(A)$, which accepts a Boolean value $A$ as input, is 1 if $A$ is true, and 0 if $A$ is false.

\subsection{Performance Criteria}
	In this paper, we use a probabilistic model and consider an average error. Assume each input $\mathbf{s} \in \mathbb{S}_{K;N}$ is with equal probability, i.e., $\mathbf{s} \sim \mathcal{U}(\mathbb{S}_{K;N})$, where $\mathcal{U}(\mathbb{A})$ means the uniform distribution in any set $\mathbb{A}$. Thus $\forall \tilde{\mathbf{s}} \in \mathbb{S}_K$, $p_{\mathbf{s}}(\tilde{\mathbf{s}}) \triangleq \text{Pr}(\mathbf{s}=\tilde{\mathbf{s}})= 1/{N \choose K}$. For a given $\mathbf{X}$, the average error probability is defined:
\begin{align}
	P^{(N)}_e(\mathbf{X})  \triangleq &\sum_{\mathbf{s}\in \mathbb{S}_{K;N}} p(\mathbf{s})\text{Pr}(d(\mathbf{s},g(\mathbf{y})) \neq 0 | \mathbf{s},\mathbf{X})\nonumber\\
=&\frac{1}{{N \choose K}}\sum_{\mathbf{s}\in \mathbb{S}_{K;N}}\sum_{\mathbf{y}}\mathbbm{1}(d(\mathbf{s},g(\mathbf{y})) \neq 0)\mathbbm{1}({\mathbf{y}=\mathbf{X}^{\top} \otimes \mathbf{s}})
\end{align}
Note that we use $p(\mathbf{s})$ instead of $p_{\mathbf{s}}(\tilde{\mathbf{s}})$ for simplification. The first term $\mathbbm{1}(d(\mathbf{s},g(\mathbf{y})) \neq 0)$ reveals the effect of decoding, and the second term $\mathbbm{1}({\mathbf{y}=\mathbf{X}^{\top} \otimes \mathbf{s}})$ the effect of channel.

    We define a number of slots $T^{(N)}_{c}$ to be achievable, if for any $T>T^{(N)}_{c}$, there exists a $N \times T$ matrix $\mathbf{X}^{(N)}$ and a decoding function $g^{(N)}(\cdot)$ for a given $N$, such that  $\displaystyle{\lim_{N \to \infty}P^{(N)}_e(\mathbf{X}^{(N)}) = 0}$. The aim is to find $T^{(N)}_{c}$, when $N\to \infty$.

\emph{Remark:}
	In group testing, the objective is to restore every user\rq{}s state, i.e., the output should be $\mathbf{z}_{g} \in \mathbb{S}_{K;N}$, and correct restoration means $\mathbf{z}_g =\mathbf{s}$. If by the definition of distortion 
\begin{align}
	d_g(\mathbf{s},\mathbf{z}_g) = \mathbbm{1}(\mathbf{z}_g=\mathbf{s}),
	\label{distortion2}
\end{align}
	the problem above is exactly a noiseless group testing
	problem. Thus the main difference between our partition problem
	and group testing problem lies in the different definitions of
	distortion functions, more importantly, lies in the different
	forms of information to transmit. Furthermore, since knowing
	$\mathcal{G}_{\mathbf{s}}$ will always induce a correct
	partition of $\mathcal{N}$ by distortion definition \eqref{distortion1}, the partition problem
	requires no more information transferred than that in the case of group
	testing. In the next section, we rigorously analyze the amount
	of the information used to solve the partition problem.

\section{Source coding}\label{sec.sourcecoding}
\begin{figure}
\centering
\includegraphics[width=3.2in]{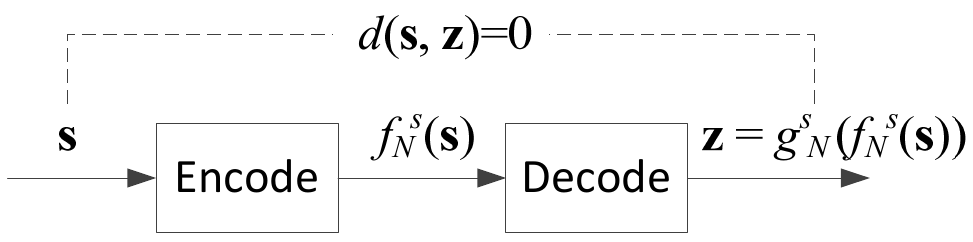}
\caption{Source coding part with distortion criterion} \label{pfig3}
\end{figure}

	In this section, we first focus on the inputs and outputs of the system without considering channel effects, i.e., a centralized source coding scheme illustrated as in Fig. \ref{pfig3}, to find the amount of information needed for describing the source with the purpose of partition. In other words, the purpose is to find a set of partitions $\mathcal{C}$ with minimum size, so that for nearly every possible $\mathbf{s} \in \mathbb{S}_{K;N}$, there is a partition $\mathbf{z} \in \mathcal{C}$ and $d(\mathbf{s},\mathbf{z})=0$. With the help of source codebook $\mathcal{C}$, for any unknown input $\mathbf{s}$, we can utilize the channel to check every partition in $\mathcal{C}$ to find the valid partition; details appear in the next section.


For group testing, the objective is to restore all states of users, if we
use a \emph{source codebook} $\mathcal{C}_g \triangleq \{\mathbf{s}_1,
\ldots, \mathbf{s}_{L^{(N)}}\}$ to represent all $\mathbf{s} \in
\mathbb{S}_{K;N}$, the size $L^{(N)}$ should be $|\mathbb{S}_{K;N}| =
{N \choose K}$. However, in the partition reservation system, for a given
$\mathbf{z} \in \mathbb{Z}_{K;N}$, there can be more than one $\mathbf{s}$
so that $d(\mathbf{s},\mathbf{z})=0$. Actually when $\mathbf{z} \in
\mathbb{Z}_{K;N}(n_1,\ldots, n_K)$, we have $\displaystyle{|\mathbb{S}_{K;N}(\mathbf{z})|
= \Pi_{k=1}^K n_k}$ number of possible active vectors so that $d(\mathbf{s}, \mathbf{z})=0$ for $\mathbf{s} \in \mathbb{S}_{K;N}(\mathbf{z})$. Thus, we can use codebook with size smaller
than $\mathbb{S}_{K;N}$ to represent the inputs. Strictly speaking,
for $\mathbf{s} \sim \mathcal{U}(\mathbb{S}_{K;N})$, if there exists a
source encoding function:
\begin{align}
	f^s_N: \mathbb{S}_{K;N} \to \{1,2,\ldots, L^{(N)}\},\nonumber
\end{align}
and a source decoding function:
\begin{align}
	g_N^s: \{1,2,\ldots, L^{(N)}\} \to \mathbb{Z}_{K;N},\nonumber
\end{align}
	so that we can map $\mathbf{s}$ to a decoding output $\mathbf{z}=g_N^s(f^s_N(\mathbf{s}))$, and the average source coding error:
\begin{align}
	P_e^{s, (N)} \triangleq \sum_{\mathbf{s} \in \mathbb{S}_{K;N}} p(\mathbf{s}) \mathbbm{1}(d(\mathbf{s},g_N^s(f^s_N(\mathbf{s})))\neq 0)
\end{align}
approaches 0 when $N \to \infty$, we will call $(L^{(N)}, f^s_N, g^s_N)$
an achievable source code sequence for the uniform source $\mathbf{s}
\sim \mathcal{U}(\mathbb{S}_{K;N})$, the range of $g^s_N(\cdot)$
is defined as the source codebook. The minimum of $\log L^{(N)}$
for all achievable source code sequences will be called the partition
information for $\mathbf{s} \sim \mathcal{U}(\mathbb{S}_{K;N})$. 

In this
section, we first compute the minimum constrained mutual information between $\mathbf{s}$
and valid partition $\mathbf{z}$, denoted by $W_N^I$,
in Lemma \ref{lem1}, and then prove the existence of an achievable source
code sequence $(L^{(N)}, f^s_N, g^s_N)$ for those $L^{(N)} > 2^{W_N^I}$
in Theorem \ref{thm1}. Note that  
when $N \to \infty$, $W_N^I$ equals to $\min I(A; U)$, and the minimum of $P_e^{s, (N)}$ equals to $Q_L(K)$, as introduced in Section II.

Constrained mutual information is always related to the rate distortion
problem \cite{cover2012elements, gallager1968information}. Thus, we first
calculate the constrained minimum mutual information for $\mathbf{s}
\sim \mathcal{U}(\mathbb{S}_{K;N})$ and valid $\mathbf{z}$, i.e.,
\begin{align}
	W_{N}^{I} \triangleq \min_{p(\mathbf{z}|\mathbf{s}) \in \mathcal{P}_{z|s}} I(\mathbf{s},\mathbf{z}) \label{mutual1}
\end{align}
where the constraint is:
\begin{align}
	\mathcal{P}_{z|s} \triangleq \left\{p(\mathbf{z}|\mathbf{s}):p(\mathbf{z}|\mathbf{s})=0, \text{if} ~d(\mathbf{s},\mathbf{z})=0\right\},
\label{ap:conditiondistr}
\end{align}
which means only valid partition $\mathbf{z}$ can be chosen for given $\mathbf{s}$. The result corresponds to Hajak \cite{1056551}, when $N \to \infty$. 

\begin{lem}
\textit{
\begin{align}	
	W_{N}^{I}\triangleq\min_{p(\mathbf{z}|\mathbf{s}) \in \mathcal{P}_{z|s}} I(\mathbf{s},\mathbf{z}) = &\log \frac{{N \choose K}}{\prod_{k=1}^{K} n^*_k}\label{originalboundsource}
\end{align}
where
\begin{align}
	(n^*_1,\ldots,n^*_K) = \arg \max_{n_k} \prod_{k=1}^{K} n_k,\nonumber\\
\text{subject to}~\sum_{k=1}^K n_k =N, ~\text{and}~ \forall k \in \mathcal{K}, n_k \geq 1.
	\label{nstarcon}
\end{align}
$W_{N}^{I}$ can be achieved by choosing
\begin{align}
	\mathbf{z}|\mathbf{s} \sim \mathcal{U}\left(\mathbb{Z}_{K;N}(n^*_1,\ldots,n^*_K) \bigcap \mathbb{Z}_{K;N}(\mathbf{s}) \right).
	\label{conditiondis}
\end{align}}
\label{lem1}
\end{lem}

Eq. \eqref{conditiondis} means for any given $\mathbf{s}$, the partition $\mathbf{z}$
should be chosen from the ``correct'' set $\mathbb{Z}_{K;N}(\mathbf{s})$
since the constraint $\mathcal{P}_{z|s}$, also we require that
$\mathbf{z} \in \mathbb{Z}_{K;N}(n^*_1,\ldots,n^*_K)$ to minimize the
mutual information, which means there are $n^*_k$ users assigned
to the group $k$. The partition $\mathbf{z}$ can be chosen uniformly from the
set satisfying these two conditions. The proof is in Appendix
\ref{appendix.a}, in which we first partition $\mathbb{Z}_{K;N}$ to
$\bigcup_{(n_1,\ldots,n_K)}\mathbb{Z}_{K;N}(n_1,\ldots,n_K)$, and then
for each set of partitions, log sum inequality is used to obtain the
lower bound. For the achievability, a direct construction of the optimal
$p(\mathbf{z} | \mathbf{s})$ is introduced by \eqref{conditiondis}.
Denote  $L^{(N)}$ as the size of a  codebook, we have 
\begin{thm}[Source coding]
\label{thm1}
\textit{There exists a codebook $\left\{\mathbf{z}_{\ell}\right\}_{\ell=1}^{L^{(N)}}$ of size $L^{(N)}$, and a source coding sequence $(L^{(N)}, f^s_N, g^s_N)$, so that for all $N$, the
average source decoding error probability is bounded by: 
\begin{align}\label{cs-known}
P^{s, (N)}_e \leq e^{-2^{\left(\log L^{(N)} - W_{N}^{I}\right)}}\nonumber
\end{align}
Thus, when $\log L^{(N)}>W_{N}^{I}$ and $\log L^{(N)}-W_{N}^{I}\overset{N \to \infty}{\longrightarrow} \infty$, sequence $(L^{(N)}, f^s_N, g^s_N)$ is achievable.
}
\end{thm}

The proof is in Appendix \ref{appendix.b}. The core of
the proof is to use random coding method to construct
the codebook $\{\mathbf{z}_{\ell}\}_{\ell=1}^{L^{(N)}}$,
in particular,	choose $\mathbf{z}_{\ell}$ i.i.d. from
$\mathcal{U}\left(\mathbb{Z}_{K;N}(n^*_1,\ldots,n^*_K)\right)$,
and show the average of $P^{s,(N)}_e$ over all possible codebooks
satisfies the bound in Theorem~\ref{thm1}, thus there must exists at least
one codebook satisfying this bound. Then by assigning the source encoding function $f_N^s(\mathbf{s}) = \arg\min_{1\leq \ell \leq L^{(N)}} d(\mathbf{s},\mathbf{z}_{\ell})$, and the source decoding function $g_N^s(\ell) = \mathbf{z}_{\ell}$, we will obtain the source coding sequence $(L^{(N)}, f^s_N, g^s_N)$ with the error probability bounded by Theorem~\ref{thm1}.

From Theorem \ref{thm1},
we can see $W_{N}^{I}$ can be used to measure the amount of asymptotic
partition information of the source. And it explicitly shows the partition
information, as well as its difference from the required information to restore
all states in further remarks.

\emph{Remark 1:}
For group testing, if we define $W_{G, N}^{I}$  as that in \eqref{mutual1}, obviously we have:
\begin{align}
W_{G, N}^{I} = &\log {N \choose K}
\end{align}
Thus $W_{N}^{I} = \log {N \choose K}-\log \left(\prod_{k=1}^{K}
n^*_k\right)$ of partition problem is smaller by a factor $\log
\left(\prod_{k=1}^{K} n^*_k\right)$ than that of group testing.
We next remark on the effect of the order of $K$ as compared
with $N$ on the achieved mutual information, as well as the error probability.

	\emph{Remark 2:} 
	First, let\rq{}s show the explicit expression of $W_{N}^{I}$. From restriction of $[n_k]_{k=1}^K$ in \eqref{nstarcon}, it is easy to see without requiring $n_k$ to be a integer, then the optimal values of $n_k$ are 
	\begin{align}
		n^*_1 = n^*_2 = \ldots = n^*_K = \frac{N}{K}.\nonumber
	\end{align}
	Thus
\begin{align}
	W_{N}^{I} \geq \log {N \choose K}-\log \left(\frac{N}{K}\right)^K \label{lowerboundsource}
\end{align}
The equality is achieved when $K$ divides $N$, and it is a good approximation when $N \gg K$. Also, we have the inequalities:
\begin{align}
	\frac{{N \choose K}}{\left(\frac{N}{K}\right)^K} \overset{(a)}{\leq} \frac{K^K}{K!} \overset{(b)}{\leq} e^K,
	\label{Kineq}
\end{align}
Equality of $(a)$ will be approximately achieved when $K \ll N$, and the equality of $(b)$ requires $1 \ll K \ll N$.

\emph{Remark 3:}
	When $K = O(N)$, e.g. $K = \eta N$ and $0<\eta<1$ is a constant, we have:
\begin{align}
	\lim_{N \to \infty} \frac{1}{N} W_{N}^{I} = &-(1-\eta)\log(1-\eta)  \label{KON}\\
	\lim_{N \to \infty} \frac{1}{N} W_{G, N}^{I} = &H(\eta) \triangleq -\eta\log \eta \label{KONG} -(1-\eta)\log(1-\eta)
\end{align}
They are obtained by a tight bound of ${N \choose K}$ derived by Wozencraft and Reiffen, see in Section 17.5 in \cite{cover2012elements}. Thus we can define an achievable source information rate $R_s$ for the partition problem (note the unit of the rate defined here is bits/user), so that for any $R \geq R_s+ \xi$, where $\xi>0$ is any constant, there exists an achievable coding sequence $(L^{(N)}=2^{N R}, f^s_N, g^s_N)$, and 
\begin{align}
	P^{s, (N)}_e \to 0, ~\text{when}~ N \to \infty
\end{align}
By Theorem \ref{thm1} and Eq. \eqref{KON}, we can see that $\displaystyle{R_s = \lim_{N \to \infty} \frac{1}{N} W_{N}^{I}}$, when $K = \eta N$, since we can always construct the achievable coding sequence of $L^{(N)}=2^{N R}$ that for all $\xi>0$, and $\forall R\geq R_s+\xi$,
\begin{align}
	P^{s, (N)}_e \leq e^{-2^{N(R-R_s)}} \to 0
\end{align}
Note that the error is doubly exponential. While for group testing, if we define $R_s^g$ similarly to $R_s$, we can see by \eqref{KON} and \eqref{KONG} that $\displaystyle{R_s^g = \lim_{N \to \infty} \frac{1}{N} W_{G, N}^{I} = R_s + (-\eta \log \eta) > R_s}$. Thus, we need higher rate to represent the states of users than to partition them.

	\emph{Remark 4:} When $K =o(N)$, $\lim_{N \to \infty} W_{N}^{I} = \log
	\frac{K^K}{K!}$. A special example is that $K$
	is a constant,  then $\lim_{N \to \infty} W_{N}^{I}$ is also a constant.  We can see the
	proposed achievable rate $R_s =0$ by \eqref{Kineq}, i.e.,
	$\frac{1}{N}W_{N}^{I} \leq \frac{K}{N} \log e \to 0$. By Theorem~\ref{thm1}, for
	any $L^{(N)} = f(N)$, where $f(N)$ is a function satisfying $f(N)
	\overset{N \to \infty}\longrightarrow \infty$, we can always
	construct a source coding sequence with codebook size $L^{(N)}
	=f(N)$, and
\begin{align}
	P^{s, (N)}_e \leq e^{-2^{\left(\log f(N)-\log \frac{K^K}{K!}\right)}} \to 0, ~\text{when}~N\to \infty
\end{align}
It can be seen that we can choose $L^{(N)}$ to be of any order of $N$ to guarantee the convergence of $P^{s, (N)}_e$, e.g., $L^{(N)}= \log \log N$. While for group testing, we should always need $L^{(N)} = {N \choose K}$ to represent the source, which can be much larger than that of partition problem. However, different choices of $f(N)$ will influence the speed of convergence, e.g., if an exponential convergence speed is required, i.e., $P^{s, (N)}_e \leq e^{-\Delta N}$ for some $\Delta >0$, there should be $L^{(N)} = O(N)$.

\section{The brute force method}\label{sec5}
Given the source codebooks randomly generated in the source coding problem, we propose a corresponding channel coding scheme. In this scheme, the channel codebook ${\bf X}$ is created by first collecting all partitions (or codewords) in the source codebook; the decoder then checks each partition (or source codeword) exhaustedly with the help of the Boolean channel. More specially, 
if the partition set $\mathcal{C}$ is given as a source codebook, and $T_0$ slots are needed to check if a partition $\mathbf{z} \in \mathcal{C}$ is a valid partition, then at most $T=T_0 \cdot |\mathcal{C}|$ slots are needed to check all partitions in $\mathcal{C}$. This is the brute force method.

\begin{figure}
\centering
\includegraphics[width=2.5in]{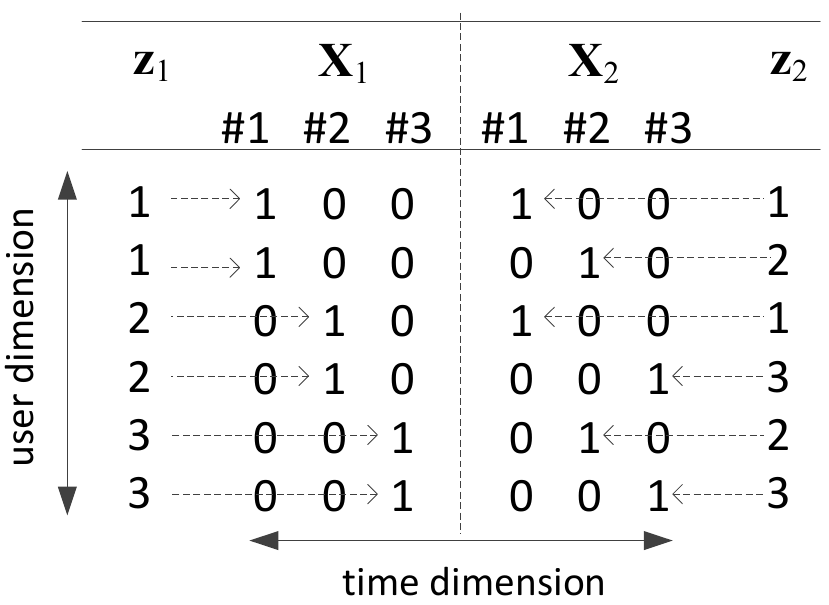}
\caption{Example of the generation of $\mathbf{X}$ in brute force method, where $N=6$, $K=3$, and source codebook of size ${L^{(N)}}=2$ is chosen.} \label{pfig4}
\end{figure}

For a given $L^{(N)}$, we can find a source codebook
$\{\mathbf{z}_{\ell}\}_{\ell=1}^{L^{(N)}}$ to represent the source under error
probability $P_e^{s,(N)}$ by Theorem \ref{thm1}. Thus if a matrix 
$\mathbf{X}$ is designed to check whether $\mathbf{z}_{\ell}$ is the correct output
one by one, the average error probability $P_e^{(N)}$ will behave the same
as $P_e^{s,(N)}$, and thus approaches zero when $\log L^{(N)} > {W^I_N}$
and $\log L^{(N)} - {W^I_N} \to \infty$. The brute force method is stated as follows:
 
\begin{enumerate}[\IEEEsetlabelwidth{12)}]
\item Source coding: For $L^{(N)}$, choose the codebook $\{\mathbf{z}_{\ell}\}_{\ell=1}^{L^{(N)}}$, and the source coding sequence $(L^{(N)}, f^s_N, g^s_N)$ based on Theorem \ref{thm1}. 
\item Joint coding: Generate $\mathbf{X}$ by $L^{(N)}$ submatrices of dimension $N \times K$, 
\begin{align}
	\mathbf{X} = [\mathbf{X}_1, \ldots, \mathbf{X}_{L^{(N)}}].\nonumber
\end{align}
Thus the dimension of $\mathbf{X}$ is $N\times T$, where $T = K {L^{(N)}}$ ($T_0 = K$ to check each possible partition). Each $\mathbf{X}_{\ell}$ is a $N \times K$ matrix, so that $\forall 1\leq i \leq N,~1\leq k \leq K$, the $(i, k)$-th element of $\mathbf{X}_{\ell}$ satisfies:
\begin{align}
	x_{\ell;i, k} = 
\begin{cases}
1, &z_{\ell; i} = k;\\
0, &\text{otherwise}
\end{cases}.\nonumber
\end{align}
See Fig. \ref{pfig4} for an example.
\item Decoding: Now the outputs are separated into ${L^{(N)}}$ blocks:
\begin{align}
	\mathbf{y} = [\mathbf{y}_1; \ldots; \mathbf{y}_{L^{(N)}}],\nonumber
\end{align}
	and
\begin{align}
	\mathbf{y}_{\ell} = \mathbf{X}_{\ell}^{\top} \otimes \mathbf{s} \nonumber
\end{align}
is a $K \times 1$ column vector. If there exists $\mathbf{y}_{\ell}
= \mathbf{1}_{K \times 1}$, where $\mathbf{1}_{K
\times 1}$ is a $K \times 1$ column vector with all components equal to 1,
then the joint decoder is $g(\mathbf{y}) =
\mathbf{z}_{\ell}$; if there exist more than one, we can select one of them,
e.g., the first one; otherwise there is decoding error. 
\end{enumerate}


Note that if $\mathbf{y}_{\ell} = \mathbf{1}_{K \times 1}$, then there
exists at least one active user in each of $k$ groups assigned by
$\mathbf{z}_{\ell}$. And since we know there are exactly $K$ active users,
only one active user is assigned in each group. Then definitely
$d(\mathbf{s},\mathbf{z}_{\ell})=0$, i.e., it is based on the following fact:

\begin{align}
	\forall i \neq j \in \mathcal{G}_{\mathbf{s}} ,z_i \neq z_j 
\iff \bigcup_{i \in \mathcal{G}_{\mathbf{s}}} \{z_i\} = \mathcal{K}.\nonumber
\end{align}

Obviously in the brute force method the number of channel uses $T_{BF}
= K{L^{(N)}}$. In addition, since in this method if there exists
$\mathbf{z}_{\ell}$ in codebook $\{\mathbf{z}_{\ell}\}_{\ell=1}^{{L^{(N)}}}$
so that $d(\mathbf{s},\mathbf{z}_{\ell})=0$, then definitely
$d(\mathbf{s},g(\mathbf{y}))=0$, the average error of the brute force
method is the same as centralized source coding. Then based on 
the analysis of centralized source coding as in Theorem \ref{thm1}, we have 
\begin{thm}[Brute force method]
	\textit{For the brute force method, if the size of centralized source codebook is ${L^{(N)}}$}, then
\begin{align}
	T_{BF} = K {L^{(N)}}, \nonumber
\end{align}
\textit{and the average error probability is}
\begin{align}
	P^{(N)}_e \leq e^{-\left(\frac{T_{BF}}{K}\middle/2^{W_{N}^{I}}\right)}
=e^{-2^{\left(\log L^{(N)} -W_{N}^{I}\right)}}\nonumber
\end{align}
\label{thm2}
\end{thm}

Although the brute force method is very simple and obviously not
optimal, it highlights some features of the partition problem. First,
if $K$ is a fixed number, then as stated in Remark 4 in the last
section, only $T_{BF} = \frac{K^{K+1}}{K!}f(N)$ is needed for the
convergence of $P^{(N)}_e$(since $P^{(N)}_e \leq e^{-f(N)}$), where
$\frac{K^{K+1}}{K!}$ is a constant and $f(N)$ is any function satisfying $\displaystyle{\lim_{N \to \infty} f(N) = \infty}$. In this case, the threshold effect of
the convergence doesn't exist as that in group testing or compressive
sensing  \cite{malyutov2013search}, and the choice of $f(N)$ is related
to the speed of convergence of $P^{(N)}_e$. However, when $K$ is large,
e.g.  when $1 \ll K \ll N$, $2^{W_{N}^{I}} \to e^K$, $T_{BF}$ should be
larger than $K e^{K}$ to guarantee the convergence of $P^{(N)}_e$,
which may be even larger than the time needed for group testing $T_G =
O(K \log N)$. It is expected since the brute force method is not optimal. In
particular, when $K$ increases, the size of centralized source codebook
increases fast, and it becomes so inefficient to check them one by one.

\section{Random coding and Reformulation as Hypergraph} \label{sec6}
	The brute force method was inspired by a centralized source coding and it works well only for small $K$. To find the achievable bound of $T$ for general case, we design the code from another way by randomly generating $\mathbf{X}$ first and then employing MAP decoding. However, to have a more amiable approach to derive an achievable rate, and to provide more insights on the internal structure of the problem in depth, a new angle from graph theory is proposed in this section, which transforms the effect of channel to a series of operations on hypergraphs. It is shown that seeking an acceptable partition is equivalent to obtaining a common strong colorable hypergraph by all users, and then coloring this hypergraph. Because we are only concerned about an achievable rate, the computational cost associated with the coloring is not counted in our framework. 
\subsection{Random coding and MAP decoding}
	Random coding is frequently used in the proof of achievability in information theory, and has been proven useful for group testing \cite{6157065}. The binary matrix $\mathbf{X}$ is generated randomly, where each element $x_{i,t} \sim \mathcal{B}(p)$ follows the i.i.d Bernoulli distribution with $p$ parameter (other distributions of $\mathbf{X}$ can also be considered, but that is out of scope of this paper). The probability of $\mathbf{X}$ is denoted by $Q(\mathbf{X})$. Then the average probability of error over the realization of $\mathbf{X}$ is given by:
\begin{align}
	P_e^{(N)} = &\sum_{\mathbf{X}}Q(\mathbf{X})P_e^{(N)}(\mathbf{X})\nonumber\\
=&\sum_{\mathbf{X}}Q(\mathbf{X})\sum_{\mathbf{s} \in \mathbb{S}_{K;N}}\sum_{\mathbf{y}} p(\mathbf{s}) p_{y|s;X}(\mathbf{y}|\mathbf{s})\mathbbm{1}(d(\mathbf{s}, g(\mathbf{y}))\neq 0)\nonumber\\
\overset{(a)}{=}&\sum_{\mathbf{X}}Q(\mathbf{X})\sum_{\mathbf{y}}p_{y|s;X}(\mathbf{y}|\mathbf{s}_0)\mathbbm{1}(d(\mathbf{s}_0, g(\mathbf{y}))\neq 0)
\label{MAPPe}
\end{align}
Since we don\rq{}t consider observation noise in this paper, 
\begin{align}
	p_{y|s;X}(\mathbf{y}|\mathbf{s}) = \mathbbm{1}\left(\mathbf{y} = \mathbf{X}\otimes \mathbf{s}\right), \nonumber
\end{align}
and equality of $(a)$ above follows from the symmetry of the generation of $\mathbf{X}$, so we can choose any particular $\mathbf{s}_0$ as input to analyze. We will choose $\mathcal{G}_{\mathbf{s}_0}=\{1,2\}$ in the rest of the paper. Since the derived achievable $T^{(N)}_c$ for random coding is of order $\log N$, we define an achievable rate $S_c$, so that for any $T$ satisfying $\frac{\log(N)}{T} \leq S_c - \xi$, where $\xi >0$ is an arbitrary constant, we have  $P^{(N)}_e \overset{N \to \infty}{\longrightarrow} 0$. Which also implies there exists a $X^*$ such that $P^{(N)}_e(\mathbf{X}^*) \overset{N \to \infty}{\longrightarrow} 0$. In this section, we will derive such a $S_c$.

	The optimal decoding method is MAP decoding, i.e., given feedback $\mathbf{y}$, choose $\mathbf{z}^* =g(\mathbf{y})$ so that $\forall \mathbf{z} \neq \mathbf{z}^* \in \mathbb{Z}_{K;N}$, the following holds
\begin{align}
	p_{z|y;X}(\mathbf{z}^*|\mathbf{y}) \geq p_{z|y;X}(\mathbf{z}|\mathbf{y}),\nonumber
\end{align}
which is equivalent to 
\begin{align}
\sum_{\mathbf{s}\in \mathbb{S}_{K;N}} \mathbbm{1}\left(d(\mathbf{s},\mathbf{z}^*) = 0\right)p_{y|s;X}(\mathbf{y}|\mathbf{s})
\geq \sum_{\mathbf{s} \in \mathbb{S}_{K;N}} \mathbbm{1}\left(d(\mathbf{s},\mathbf{z}) = 0\right)p_{y|s;X}(\mathbf{y}|\mathbf{s}) \label{MAPdecode}
\end{align}
If there is more than one $\mathbf{z}^*$ with the maximum value, choose any one. Note that here we search all possible $\mathbf{z}^*$ in all possible $\mathbf{z} \in \mathbb{Z}_{K;N}$; however, considering the source coding results, we can just search $\mathbf{z}^* \in \mathbb{Z}_{K;N}(n_1^*, \ldots, n_K^*)$ without loss of generality. 

	As seen in the definition of MAP decoding, to find MAP of $\mathbf{z}$, we should count all $\mathbf{s} \in \mathbb{S}(\mathbf{z})$ satisfying $\mathbf{y}=\mathbf{X}\otimes \mathbf{s}$. While many $\mathbf{s} \in \mathbb{S}(\mathbf{z})$ has common active users, so $\mathbbm{1}(\mathbf{y}=\mathbf{X}\otimes \mathbf{s})$ are correlated for different $\mathbf{s}$ sharing parts of common active users. Thus, it is extremely difficult to compare the posterior probability of different $\mathbf{z}$. The obstacle arises because in MAP decoding, few inherent structures of the problem are found and utilized. To further reveal this inherent problem structure, a novel formulation from the perspective of hypergraph is proposed in the next section, which proves to be helpful in reducing complexity of performance analysis.

\subsection{Reformulation as Hypergraph}
\begin{figure}
\centering
\includegraphics[width=3.2in]{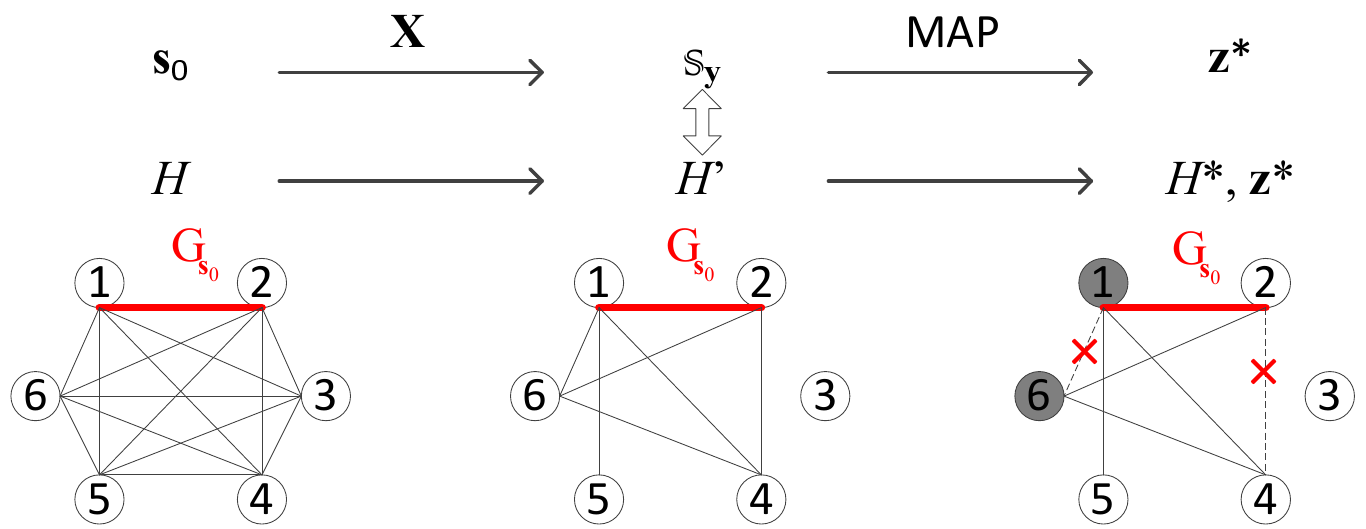}
\caption{Reformulation from hypergraph} \label{pfig6}
\end{figure}

	The process of random coding can be illustrated in the upper part of Fig. \ref{pfig6}. For an input $\mathbf{s}_0$, the channel output $\displaystyle{\mathbf{y}=\bigvee_{i \in \mathcal{G}_{\mathbf{s}_0}} \mathbf{x}_i}$ is observed, and then a candidate subset of $\mathbb{S}_{K;N}$ that is capable of generating $\mathbf{y}$ can be inferred:
\begin{align}
	\mathbb{S}_{\mathbf{y}} = \left\{\mathbf{s} \in \mathbb{S}_{K;N}: \mathbf{y}=\mathbf{X} \otimes \mathbf{s}\right\} \nonumber
\end{align} 
MAP decoder tries to find $\mathbf{z}^*$ such that there is the largest number of $\mathbf{s} \in \mathbb{S}_{\mathbf{y}}$ satisfying $d(\mathbf{z}^*, \mathbf{s})=0$.

	This process can be illustrated from the perspective of hypergraphs, as shown in Fig. \ref{pfig6}.


\begin{enumerate}
	\item Source: Since all real sources $\mathbf{s}_0 \in \mathbb{S}_{K;N}$ are equiprobable, a complete $K$-uniform hypergraph $H(V(H), E(H))$ can be used to express the knowledge of the source before observation, where the set of nodes $V(H) = \mathcal{N}$ represents $N$ users, and the set of hyper-edges $E(H) = \{e \subseteq V(H) : |e| = K\}$ represents all possible inputs  \cite{west2001introduction, berge1973graphs}. It means every hyper-edge in $H$ could be $\mathbf{s}_0$. Actually the real input is just an edge $\mathcal{G}_{\mathbf{s}_0} \in E(H)$, the objective of group testing is to find exactly this edge to obtain every user\rq{}s state; while for partition reservation system, the objective is to separate each vertex of $\mathcal{G}_{\mathbf{s}_0}$. 
	\item Transmission and observation: the transmission and corresponding observation can be seen as a series of edge deleting operations on the hypergraphs. Because after observing each feedback $y_t$, $1 \leq t \leq T$, some $\mathbf{s}$ could be determined to be not possible, and the candidate set  $\mathbb{S}_{\mathbf{y}}$ could shrink. A sub-hypergraph $H\rq{}(V(H\rq{}),E(H\rq{})) \subseteq H(V(H),E(H))$ is used to denote the candidate set $\mathbb{S}_{\mathbf{y}}$ after observing the feedback $\mathbf{y}$. Note that we consider the node set $V(H\rq{})=V(H)$ to be invariant, but actually there will be many isolated nodes in $V(H\rq{})$ with zero degree. The details of the operations will be shown in next subsection. Note that for the considered noiseless case, we always have $\mathbf{s}_0 \in \mathbb{S}_{\mathbf{y}}$, so $\mathcal{G}_{\mathbf{s}_0} \in H\rq{}$.
	\item Partition: Finally, the partition $\mathbf{z}^*$ should be decided by observing $H\rq{}$. First, we introduce the concept of strong coloring. A strong coloring of a hypergraph $H$ is a map $\Psi : V(H) \to \mathbb{N}^+$, such that for any vertices $u, v \in e$ for some $e \in E(H)$,   $\Psi(u) \neq \Psi(v)$. The value of $\Psi(u)$ is called the color of node $u$. In other words, all vertices of any edge should have different colors.The corresponding strong chromatic number $\chi_s(H)$ is the least number of colors so that $H$ has a proper strong coloring \cite{agnarsson2005strong}. Obviously for a $K$-uniform hypergraph, $\chi_s(H) \geq K$. We called a strong coloring with $K$ colors to be $K$-strong coloring. If $z_{i}^*$ is viewed as a color of node $i$, actually $\mathbf{z}^* \in \mathbb{Z}_{K;N}$ gives a coloring mapping of $V(H)$ with $K$ colors. 

For MAP decoding in \eqref{MAPdecode}, the method of finding $\mathbf{z}^*$ from $\mathbb{S}_{\mathbf{y}}$ is equivalent to finding a hypergraph $H^*(V(H^*), E(H^*)) \subseteq H\rq{}(V(H), E(H\rq{}))$, such that $\chi_s(H^*)=K$, i.e., $H^*$ is $K$-strong colorable, and the number of deleted edges $|E(H\rq{}) \setminus E(H^*)|$ is minimum. Then the output $\mathbf{z}^*$ can be any strong coloring of $H^*$. 
\end{enumerate}

	From the prospective of hypergraph, the process can be represented as $H \to H\rq{} \to (H^*, \mathbf{z}^*)$, corresponding to the expression from vectors $\mathbf{s}_0 \to \mathbb{S}_{\mathbf{y}} \to \mathbf{z}^*$. The process is shown in Fig. \ref{pfig6} through an example of $N=6$, $K=2$. Note that the hypergraph becomes a graph when $K=2$. Compared with group testing, whose objective is to obtain $H^*=H\rq{}$ with only one edge $\mathcal{G}_{\mathbf{s}_0}$ by deleting edges through transmissions and observations, our partition problem allows $H\rq{}$ and $H^*$ to have more edges, so less effort is needed to delete edges, which is translated to higher achievable rate than that of the group testing problem. We can see $\mathbf{z}^*$ is correct iff $\mathcal{G}_{\mathbf{s}_0} \in E(H^*)$ and $H^*$ is $K$-strong colorable, we will use this equivalent condition to judge if the decoding is correct in the analysis. 

From the view of the algorithm, first  all users obtain a common \lq\lq{}good\rq\rq{} $H\rq{}$ which will lead to a correct partition. Second, we obtain $H^*$ and choose a common consistent $\mathbf{z}^*$. The second step, to obtain $H^*$ by deleting the minimum number of edges from $H\rq{}$  and find the $K$-strong coloring \cite{agnarsson2005strong}, does not influence the transmission time needed to arrive at a \lq\lq{}good\rq\rq{} $H\rq{}$. This is because once all users have the same copy of $H^{\prime}$, the remaining computation, including removal of edges  and coloring, can be done locally without further expending communication resources. A further explanation of operations of the deleting edges  is introduced in the next subsection.

\subsection{Reduction step: obtaining $H\rq{}$ from $H$}\label{effectoftests}
\begin{figure}
\centering
\includegraphics[width=3.2in]{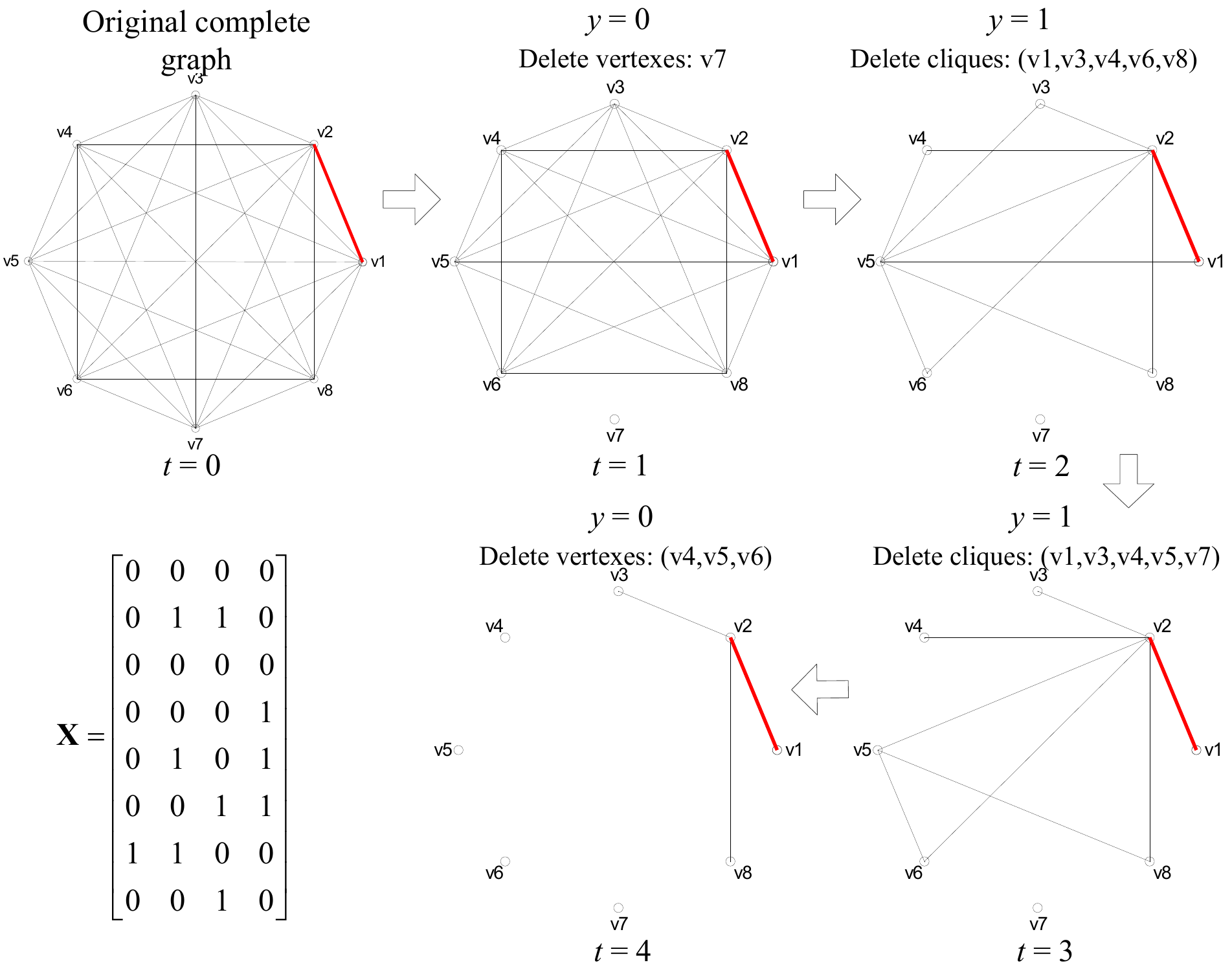}
\caption{Example of $\mathbf{X}$ effects on the operations of a graph. (Here $N=8$, $K=2$, $T=4$, and $\mathcal{G}_{\mathbf{s}_0} = \{1,2\}$)} \label{pfig7}
\end{figure}

	The effect of transmissions and observation using matrix $\mathbf{X}$ can be summarized in two operations: deleting vertex and deleting clique. Assume at time $t$, the set of active users transmitting $1$ is:
\begin{align}
	\mathcal{G}_{\mathbf{X}}(t) = \{i \in \mathcal{N}: x_{i,t} = 1\}.
\end{align}
	The operation at time $t$ can be classified based on the feedback $y_t$:
\begin{enumerate}
	\item If $y_t=0$, which means any users in $\mathcal{G}_{\mathbf{X}}(t)$ should not be active users, so these vertices should be deleted, i.e., all edges containing these vertices should be deleted. 
	\item If $y_t=1$, which means at least one active user is transmitting 1 at time $t$, so any edge completely generated by vertices in $\mathcal{N}\setminus \mathcal{G}_{\mathbf{X}}(t)$ should be deleted. Otherwise if these edges are actually the $\mathcal{G}_{\mathbf{s}_0}$, there will be no active users in $\mathcal{G}_{\mathbf{X}}(t)$ and $y_t=0$. In fact, it is equivalent to deleting all $K$ uniform hyper-cliques generated by $\mathcal{N} \setminus \mathcal{G}_{\mathbf{X}}(t)$.
\end{enumerate}
	The two effects are illustrated by an example in Fig. \ref{pfig7} using a graph with sequence of nodes removed and clique removing operations. There are $8$ users and $4$ slots are used for transmission. We can see the edges removing process starting from a complete hypergraph at $t=0$, to a graph of only 3 edges at time $t=4$. At $t=1, 4$, $y_t=0$, the corresponding vertices are removed, while at time $t=2,3$ cliques are removed.

	Now it is clear that our problem can be viewed as a $K$-strong hypergraph coloring problem, and the objective is to schedule a series of edge removing operations efficiently to construct such a hypergraph so that all $K$ active users could be assigned a unique color (or transmission order).  In next section, a special case of $K=2$ is solved; even in this simple case, the problem is nontrivial.


\section{Random coding of $K=2$}\label{sec7}
	For $K=2$, two sub-optimal decoding methods inspired by MAP decoding are proposed to further simplify the calculation.

\subsection{Two simplified decoding methods}
	In MAP decoding, the decoder will find a $K$-strong colorable graph $H^*$ from $H\rq{}$ by deleting the minimum number of edges, and the decoding result is correct if $\mathcal{G}_{\mathbf{s}_0} \in H^*$. For $K=2$, hypergraph $H^*$ being $2$-strong colorable is equivalent to $H^*$ being bipartite, or equivalently, having \emph{no odd cycles}. 	Further, assume $\mathcal{G}_{\mathbf{s}_0} = \{1,2\}$, odd cycles can be classified into three kinds: 
\begin{enumerate}
	\item Containing a cycle with vertices 1 and 2, the cycle may or may not containing edge (1,2);
	\item Containing one of the vertices 1 and 2;
	\item Containing neither vertex 1 or 2.
\end{enumerate}
Denote the odd cycles containing edge $(1,2)$ by \lq\lq{}1-odd cycles\rq\rq{}. Since $(1,2)$ always exists in $H\rq{}$ due to the noiseless channel, it is easy to see $H\rq{}$ contains no first kind of cycles iff
$H\rq{}$ contains no 1-odd cycles. Thus in the rest of paper, we just consider the existence of $1$-odd cycles and the other two kinds of odd cycles (sometimes for simplification of notation, we also call $1$-odd cycle \emph{the type-1 odd cycle}). We can assert that if there is no 1-odd cycles in $H\rq{}$, the decoding result will be surely correct. The reason is that for MAP decoding, it breaks all odd cycles in $H\rq{}$ to get $H^*$ by deleting least edges. If there is no 1-odd cycle in $H\rq{}$, set $\mathcal{G}_{\mathbf{s}_0}$ will not be deleted during this process. Thus, $\mathcal{G}_{\mathbf{s}_0} \in H^*$, which implies the correct decoding. Thus, we have
\begin{align}
	P_e^{(N)} \leq &P_e^{1-odd} \triangleq \sum_{\mathbf{X}} Q(\mathbf{X})\text{Pr}(\text{$H\rq{}$ contains 1-odd cycles}|\mathbf{X}, \mathbf{s}_0)\\
\leq &P_e^{odd} \triangleq \sum_{\mathbf{X}} Q(\mathbf{X})\text{Pr}(\text{$H\rq{}$ contains odd cycles}|\mathbf{X}, \mathbf{s}_0)
\end{align}

	 In the following, $P_e^{odd}$ and $P_e^{1-odd}$ are both upperbounded by their respective union bounds, and it is shown their upperbounds are nearly the same when $N \to \infty$, which points to the possibility of using a suboptimal decoding to advantage: when $H\rq{}$ is 2-colorable, find any $\mathbf{z}$ consistent with it; otherwise announce an error. The reason is if MAP coding is used, it is necessary to obtain $H^*$ by deleting minimum edges of $H\rq{}$, which is a NP hard problem\cite{yannakakis1978node}; however, it is easy to judge whether $H\rq{}$ is a bipartite graph in linear steps of $N$. So while the suboptimal decoding method needs more channel use,  it is easier to compute.

\subsection{Main result: Achievable bound of $T$ for $K=2$ case}
	To upperbound $P^{(N)}_e$ by $P_e^{1-odd}$, denote:
\begin{align}
	C(p) = -(1-(1-p)^2)\log \varphi(p) -(1-p)^2\log (1-p)
\end{align}
where
\begin{align}
	\varphi(p) = \frac{p+\sqrt{4p-3p^2}}{2} \label{phipthm}.
\end{align}
We have the following lemma:
\begin{lem}\label{thm3}
	\textit{For $K=2$ case, if for any constant $\xi>0$ such that $\frac{\log N}{T} \leq S_{c}-\xi $, and $S_c \triangleq \displaystyle{\max_{0\leq p \leq 1} C(p)}$, we have $P^{(N)}_e \leq P_e^{1-odd}  \xrightarrow{N \to \infty} 0$.} 
\end{lem}

	And similarly, by bounding $P^{(N)}_e$ by $P_e^{odd}$, we have the following theorem:
\begin{thm}
	\textit{For $K=2$ case, for any constant $\xi>0$ that $\frac{\log N}{T} \leq S_c-\xi$, we will have $P^{(N)}_e \leq P_e^{odd} \xrightarrow{N \to \infty} 0$.}
\label{prop2}
\end{thm}

	The proofs of Lemma \ref{thm3} and Theorem \ref{prop2} are given in Appendix \ref{appendix.c}. Actually if the elements of $\mathbf{X}$ are generated i.i.d. by Bernoulli distribution of parameter $p$, we will have $P_e^{1-odd}$ and $P_e^{odd}$ approaches 0 if $\frac{\log N}{T} \leq C(p)-\xi$, thus $\displaystyle{S_c = \max_p C(p)}$. We can see the achievable bounds are both $S_c$ for two methods to make $P^{(N)}_e \to 0$. The main idea in the proof is to calculate the probability of existence of a particular odd cycle in $H\rq{}$, and the calculation is similar for all three kinds of odd cycles. As a key factor in the result, $\varphi(p)$ in \eqref{phipthm} is actually a factor of the solution of the extended Fibonacci numbers, which reveals some interesting structure in the partition problem. 

	A sketch of the proof of Lemma \ref{thm3} is given as below, it is the same for Theorem \ref{prop2}:
\begin{enumerate}
\item Consider the problem conditioning on $[\mathbf{x}_1,\mathbf{x}_2]$ in a strong typical set $\mathcal{A}_{\epsilon}^{(T)}$; this will make the algebra easier. Assume the probability of existence of a particular 1-odd cycle of $M$ vertices in $H\rq{}$ to be $P_{e;M}$; there are ${N-2 \choose M-2} (M-2)! \leq N^{M-2}$ such odd cycles and all of them are equiprobable. Thus, 
\begin{align}
	P_e^{1-odd} \leq \sum_{M\geq 3, M~\text{is odd}} 2^{(M-2)\log N}P_{e;M} + \text{Pr}([\mathbf{x}_1, \mathbf{x}_2] \in \mathcal{A}_{\epsilon}^{(T)}) \label{exp1}
\end{align}
Since $\text{Pr}([\mathbf{x}_1, \mathbf{x}_2] \in \mathcal{A}_{\epsilon}^{(T)}) \leq 2^{-q(p,\epsilon) T} \overset{T \to \infty}{\longrightarrow} 0$, where $q(p,\epsilon)$ is some constant, according to the properties of strong typical set\cite{csiszar2011information}. We will show that $P_{e;M} \leq 2^{-(M-2)C(p)T}$. Thus when $\log N < (C(p)-\xi) T$, $2^{(M-2)\log N}\times P_{e;M} \leq 2^{-(M-2)\xi}$, which means the $P_e^{1-odd}\to 0$. Note that we can also see the $P_e^{1-odd}$ goes to 0 with an exponential speed with $T$, and thus, polynomial speed with $N$, i.e., $P_e^{1-odd} \leq 2^{-\Delta_1 T} = \frac{1}{N^{\Delta_2}}$, where $\Delta_1$ and $\Delta_2$ are constant.

\item Divide $T$ slots  into four parts $\mathcal{T}_{u,v} = \{t: (x_{1,t},x_{2,t}) = (u,v)\}$, for four different $(u, v) \in \{0,1\}^2$, according to the codewords of the real input $[\mathbf{x}_1, \mathbf{x}_2]$. In the strong typical set, we just need to consider when $|\mathcal{T}_{u,v}| \approx p_x(u)p_x(v)T$, where $p_x(u) \triangleq p\mathbbm{1}(u=1)+ (1-p)\mathbbm{1}(u=0)$ is the probability distribution of Bernoulli variable. And due to symmetry and independence of the generation of $\mathbf{X}$, for any $(u,v)$, we just need to consider any slot $t$  $\mathcal{T}_{u,v}$.
\item At $t \in \mathcal{T}_{u,v}$, denote $\mu_{u,v;M}$ to be the probability that the considered 1-odd cycle of length $M$ won\rq{}t be deleted by the operations, then
\begin{align}
	P_{e;M} = \displaystyle{\Pi_{u,v} \left(\mu_{u,v;M}\right)^{|\mathcal{T}_{u,v}|}\approx \Pi_{u,v} \left(\mu_{u,v;M}\right)^{p_x(u)p_x(v)T}}\label{exp2}
\end{align}
\item We have shown that for all $t$ where $y_t = 0$, i.e., $\displaystyle{t \in \bigcup_{(u,v)\neq (0,0)}\mathcal{T}_{u,v}}$, $\mu_{u,v;M} = (1-p)^{M-2}$, and the exponent $p_x(0)p_x(0) = (1-p)^2$, thus 
\begin{align}
\displaystyle{\left(\mu_{0,0;M}\right)^{p_x(0)p_x(0)T} = (1-p)^{(M-2)(1-p)^2T} = 2^{(M-2)T(1-p)^2\log (1-p)}} \label{11}
\end{align}

While for $y_t = 1$, i.e., $t \in \mathcal{T}_{u,v}$, $(u,v)\neq (0,0)$, we have shown $\mu_{u,v;M} = \varphi^{M-2}(p)$, and $\sum_{(u,v)\neq (0,0) } p_x(u)p_x(v)=1-(1-p)^2$. Thus, 
\begin{align}
\Pi_{(u,v)\neq 0}\left(\mu_{u,v;M}\right)^{p_x(u)p_x(v)T} = (1-p)^{(M-2)(1-(1-p)^2)T} = 2^{(M-2)T(1-(1-p))^2\log \varphi(p)} \label{22}
\end{align}
 Then, combining \eqref{11} and \eqref{22}, we obtain $P_{e;M} \leq 2^{-(M-2)C(p)T}$, which completes the proof.
\end{enumerate}

	We can provide an intuitive explanation. The result in Lemma \ref{thm3} can be expressed as
\begin{align}
	\forall p, ~T > \frac{\log N^{M-2}}{(M-2) C(p)}
\end{align}
Intuitively the problem can be stated as that we have at most $N^{M-2}$ 1-odd cycles of length $M$, and after determining(or eliminating) all of them, the error probability becomes 0. Thus, $\log N^{M-2}$ can be seen as an upperbound on source information, which describes the uncertainty of 1-odd cycles; ${(M-2)} C(p)$ can be seen as the information transmitting rate of the channel, which represents the speed of eliminating the uncertainty of odd cycles with $M$ vertices. 

\begin{figure}
\centering
\includegraphics[width=1.8in]{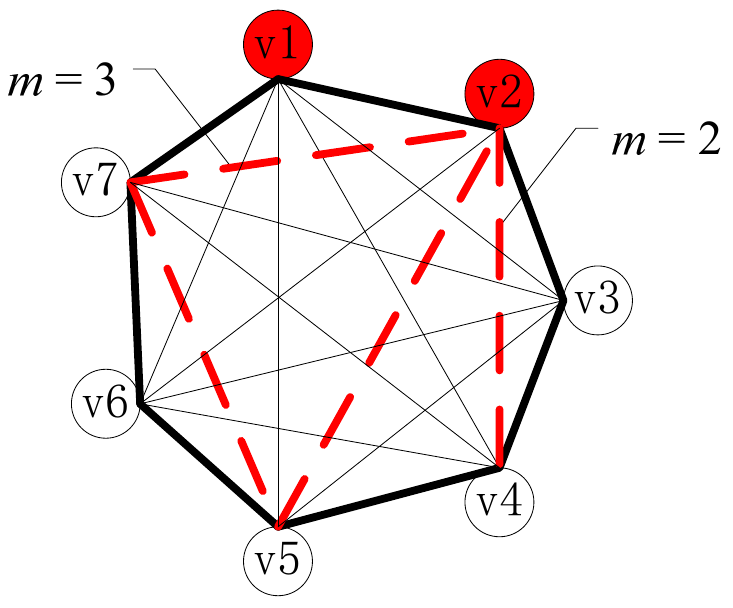}
\caption{Random clique deleting while keeping an particular odd cycle at a particular $t$ such that $\mathbf{y}(t)=1$. (Here the size of the odd cycle is $M=7$, $K=2$, only the cliques of size 2 (edge) or 3 (triangle) consisted with non consecutive vertices can be deleted, as shown by $(2,4)$ and $(2,5,7)$ for example.)} \label{pfig8}
\end{figure}

	To further explain the meaning of $(M-2)C(p)$, we should use the effect of $\mathbf{X}$ on hypergraph stated in Section \ref{effectoftests}. If a given 1-odd cycle $H_{e;M}$ of $M$ vertices exists in $H\rq{}$, in $1\leq t \leq T$, none of the $M$ vertices, or the cliques containing the edges of $H_{e;M}$ can be deleted. See an example in Fig. \ref{pfig8}, where $H_{e;M}$ is the outer boundary. It won\rq{}t be removed if all the nodes are maintained; and the cliques to be deleted should not contain the consecutive vertices on the outer boundary.

 	Since at any test $t$, vertices are deleted if $y_t=0$; the probability of this happening is $(1-p)^2$. For a particular $t$ when $y_t = 0$, the vertex of an inactive vertex $i$ is deleted only if $x_{i,t} = 1$, so the probability that all $M$ vertices are maintained at time $t$ is $\mu_{0,0;M} = (1-p)^{M-2}$. 

	On the other hand, all the edges of the odd cycle $H_{e;M}$ can't be deleted by the clique deleting operation. At any slot $t$ so that $y_t = 1$, whose probability is $1-(1-p)^2$, there are 3 different cases $(x_{1,t},x_{2,t}) = (u,v), (u,v) \neq (0,0)$, their analysis is similar, let us just consider $(x_{1,t},x_{2,t})=(1,1)$. Assume $H_{e;M} = (1,2,i_1,\ldots,i_{M-2})$, so at any slot $t$, the probability that $H_{e;M}$ is not removed by deleting cliques can be derived:
\begin{align}
	\mu_{1,1;M}=&1- \text{Pr}(H_{e;M}~\text{is removed at slot $t$}|t \in \mathcal{T}_{1,1})\nonumber\\
=&1- \text{Pr}(\exists  w \in\{1,\ldots,M-3\}, 
	(x_{i_{w}}(t),x_{i_{w+1}}(t))=(0,0)) \nonumber\\
\overset{(a)}{=} &\frac{1}{p} F(M,p)
\leq \varphi(p)^{M-2} \label{ineqt}
\end{align}
The derivation of $(a)$ is seen in Appendix \ref{appendix.c}, and
\begin{align}
F(k,p) = &\sum_{j=0}^{\lfloor \frac{k-1}{2} \rfloor} {k-1-j \choose j} p^{k-1-j}(1-p)^j\\
	=&\frac{\varphi(p)^k - \psi(p)^k}{\varphi(p) - \psi(p)},
\end{align}
\begin{align}
	\textit{where}\quad \quad \varphi(p) = \frac{p+\sqrt{4p-3p^2}}{2}; \quad \psi(p) = \frac{p-\sqrt{4p-3p^2}}{2}
\end{align}
is the solution of a generalized Fibonacci sequence \cite{nguyen2012generalized}. Actually $\frac{1}{p}F(k+2,p)$ is the probability that there are no consecutive 0s in a $p$-Bernoulli sequence of length $k$. This feature of Fibonacci sequences has also been used in generating codes without consecutive 1s, known as Fibonacci coding. The other $\mu_{1,0;M}$ and $\mu_{0,1;M}$ can be derived similarly. Thus, we can see $(M-2)C(p)$ is well explained as the rate of deleting vertices or cliques for an 1-odd cycle with $M$ vertices from above.

	Now it is clear that Lemma \ref{lem1} and Theorem \ref{thm3} above have revealed the internal structure of the partition problem. The partition information is related to odd cycles, and $\mathbf{X}$ is constructed to destroy the odd cycles by deleting vertices or cliques. The Fibonacci structure emerges since it is related to considering consecutive 0s in Bernoulli sequences, which may be a key factor in partition problem and could be extended to more general cases with $K>2$. In the next section, the efficiency is compared with random coding based group testing approach.


\section{Comparison}\label{sec8}
	As stated in the introduction, our partition reservation has close relation to direct transmission and group testing. Since the average error considered in direct transmission system is not the same as the definition used in this paper, we just compare with the group testing. 

\begin{figure}
\centering
\includegraphics[width=3.5in]{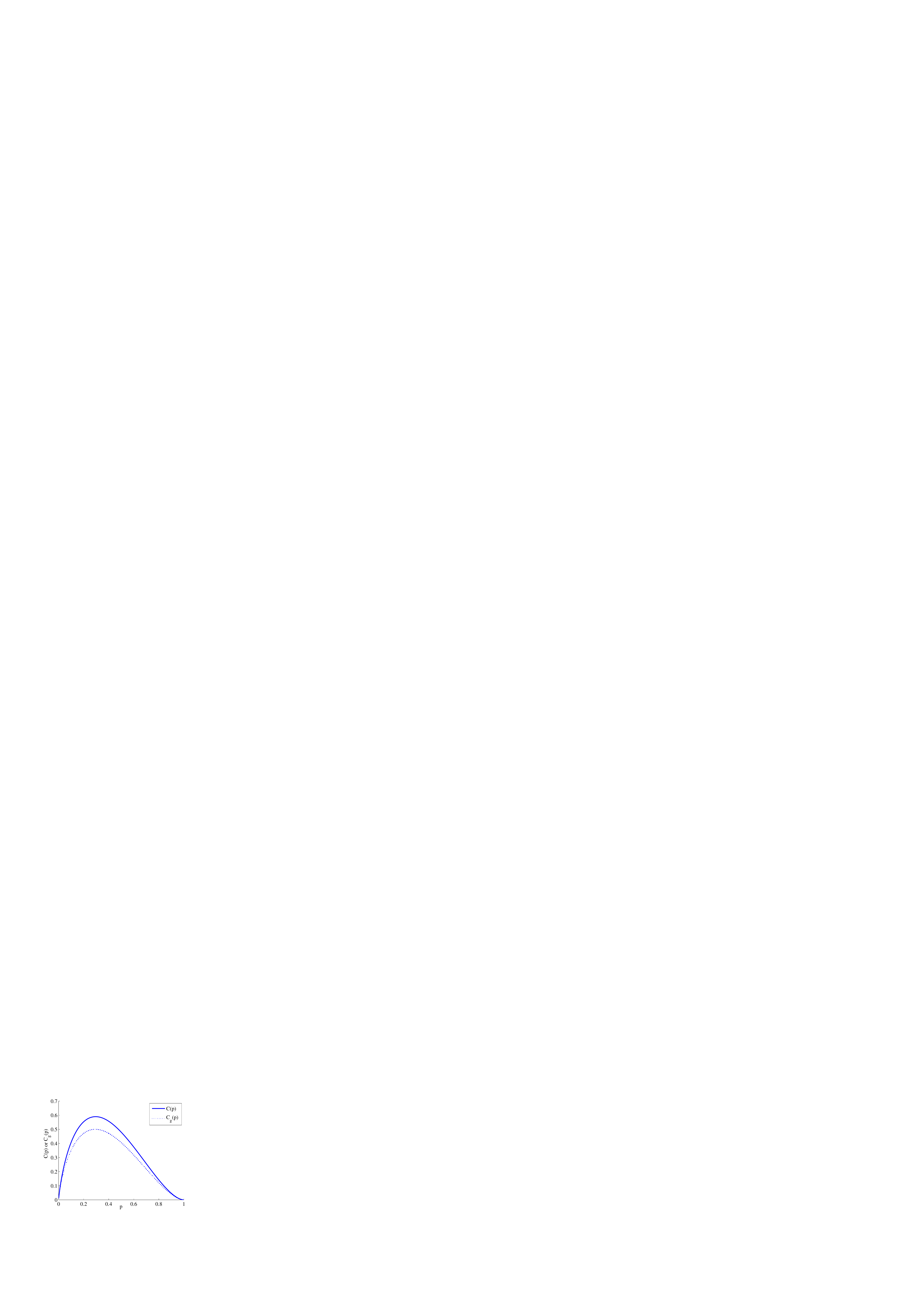}
\caption{Compare $C(p)$ with $C_g(p)$. } \label{pfig11}
\end{figure}

     Atia and Saligrama have proved the achievable rate in Theorem III. 1 in \cite{6157065} for group testing with random coding, which shows that if for any $\xi>0$ and $\frac{\log N}{T} \leq S_{cg}-\xi$, $S_{cg} \triangleq \displaystyle{\max_{p} C_g(p)}$, where
\begin{align}
	C_g(p) = \min \left\{(1-p) H(p), \frac{1}{2}H((1-p)^2)\right\}\nonumber
\end{align}
the average error probability $P_e^{(N)} \to 0$ (Also, $S_{cg}$ is shown to be a capacity in Theorem~{IV. 1} in \cite{6157065}). From Fig.~\ref{pfig11}, we can see $C_g(p) <C(p)$ for any $0<p<1$. In particular, $S_{cg} = \max_p C(p)=0.5 <0.5896 =\max_p C_g(p)=S_{c}$; i.e., our achievable rate is always larger than the capacity of group testing with random coding in the noiseless case, when using random coding. Further, the diminishing speeds of group testing and partition reservation are both polynomial, i.e., $P_e^{(N)} \leq \frac{1}{N^\Delta}$ for some constant $\Delta>0$.

	Compared with a brute force method, as shown in Section \ref{sec5}, when $K=2$, if $T_{BF} = \frac{K^{K+1}}{K!}f(N)$, where $f(N)$ is an arbitrary function satisfying $f(N) \xrightarrow{N \to \infty} \infty$, we will have $P_e^{(N)} \leq e^{-f(N)}\xrightarrow{N \to \infty} 0$. This means that the threshold effect of the convergence doesn't exist as that in group testing or compressive sensing  \cite{malyutov2013search}, i.e., $T$ should be of order $\log N$. However, the choice of $f(N)$ will influence the diminishing speed. If we require that the convergence speed of the brute force method to be polynomial,  $f(N) = O(\log N)$ and thus $T_{BF}=O(\log N)$, which is of the same order as partition and group testing. 

	The random coding method is not as efficient as the brute force method when $K=2$ on two counts: first, the result derived by the random coding method has a threshold effect for the convergence of $P_e^{(N)}$, namely, $P_e^{(N)} \to 0$ only when $T =c \log N$; while the result derived by brute force method does not have such constraint.  Second, even when $T$ is constrained to be $T = c \log N$, and $c \geq 1/C(p)$, the upperbounds for error probability with the random coding approach satisfies $P^{(N)}_{e;1} \leq N^{-\Delta_1}$, and  $P^{(N)}_{e;2} \leq N^{-\Delta_2}$ for the brute-force method, where $\Delta_1 < \Delta_2$. This means that the error probability under the brute-force approach decays faster than that using the random coding method. The reason is because intuitively, in the brute force approach, we actually encode the coloring information in the codebook, which is not the case for random coding. The main point is if we use the source codebook to construct channel codebook as done in the brute force approach, we have to deliver the associated coloring information. While for the random coding approach, we actually only care about sending information enough for the nodes to form a 2-colorable graph. However, random coding approach shows the internal structure of the problem, and the possibility to attain consistent partition for generalized $K>2$.


\section{Conclusion}\label{sec9}
    In this paper, a new partition reservation problem is formulated to focus on the coordination overhead in the conflict resolution multi-access problem. The partition information, which is related to the relationships between active users is calculated, and two codebook design methods, source coding based and random coding are proposed and analyzed to estimate the achievable bound of partitioning overhead in non-adaptive $(0,1)$-channel. The formulation using hypergraph and its coloring reveals the partition information and fundamental properties of the decentralized Boolean channel, and further uncovers the Fibonacci structure for a non-trivial $K=2$ case. The comparison with group testing shows the uniqueness of the partitioning problem, and sheds light on the future designs. In this paper, we just provide achievable rate for simple $K=2$ cases, the converse bound and more general $K>2$ cases are our ongoing works. In addition, we are working on cases with noise present in multiple access channels, which requires different machinery in the corresponding achievability analysis for the random coding approach \cite{wsh3}. 

\appendices

\section{Proof of Lemma \ref{lem1}}\label{appendix.a}
\begin{IEEEproof}
	The following derivation is subject to $p(\mathbf{z}|\mathbf{s}) \in \mathcal{P}_{z|s}$:
\begin{align}
	I(\mathbf{s};\mathbf{z}_{\mathbf{s}}) 
= &H(\mathbf{s})-H(\mathbf{s} \mid \mathbf{z})\nonumber\\
= &H(\mathbf{s}) + \sum_{\mathbf{z} \in \mathbb{Z}_{K;N}}\sum_{\mathbf{s} \in \mathbb{S}_{K;N}(\mathbf{z})}p(\mathbf{z} \mid \mathbf{s}) p_{s}(\mathbf{s})
\log \frac{p(\mathbf{z} \mid \mathbf{s})p_{s}(\mathbf{s})}{\sum_{\tilde{\mathbf{s}} \in \mathbb{S}_{K;N}(\mathbf{z})}p(\mathbf{z} \mid \tilde{\mathbf{s}})p_{s}(\tilde{\mathbf{s}})}\nonumber\\
\overset{(a)}{=} &\log {N \choose K} + p_{s}(\mathbf{s})\sum_{\mathbf{z} \in \mathbb{Z}_{K;N}}\sum_{\mathbf{s} \in \mathbb{S}_{K;N}(\mathbf{z})}p(\mathbf{z} \mid \mathbf{s})
\log \frac{p(\mathbf{z} \mid \mathbf{s})}{\sum_{\tilde{\mathbf{s}} \in \mathbb{S}_{K;N}(\mathbf{z})}p(\mathbf{z} \mid \tilde{\mathbf{s}})}\nonumber\\
\overset{(b)}{=} &\log {N \choose K} + p_{s}(\mathbf{s})\sum_{(n_1,\ldots,n_K)}\sum_{\mathbf{z} \in \mathbb{Z}_{K;N}\{n_1,\ldots,n_K\}}\sum_{\mathbf{s} \in \mathbb{S}_{K;N}(\mathbf{z})}p(\mathbf{z} \mid \mathbf{s})
\log \frac{p(\mathbf{z} \mid \mathbf{s})}{\sum_{\tilde{\mathbf{s}} \in \mathbb{S}_{K;N}(\mathbf{z})}p(\mathbf{z} \mid \tilde{\mathbf{s}})}\nonumber\\
\overset{(c)}{\geq} &\log {N \choose K} + p_{s}(\mathbf{s})\sum_{(n_1,\ldots,n_K)}\sum_{\mathbf{z} \in \mathbb{Z}_{K;N}\{n_1,\ldots,n_K\}}\left(\sum_{\mathbf{s} \in \mathbb{S}_{K;N}(\mathbf{z})}p(\mathbf{z} \mid \mathbf{s})\right) \times\nonumber\\
&\log \frac{\left[\sum_{\mathbf{s} \in \mathbb{S}_{K;N}(\mathbf{z})}p(\mathbf{z} \mid \mathbf{s})\right]}
{\sum_{\mathbf{s} \in \mathbb{S}_{K;N}(\mathbf{z})}\left[\sum_{\tilde{\mathbf{s}} \in \mathbb{S}_{K;N}(\mathbf{z})}p(\mathbf{z} \mid \tilde{\mathbf{s}})\right]}\nonumber\\
{=} &\log {N \choose K} - p_{s}(\mathbf{s})\sum_{(n_1,\ldots,n_K)}\sum_{\mathbf{z} \in \mathbb{Z}_{K;N}(n_1,\ldots,n_K)}\left(\sum_{\mathbf{s} \in \mathbb{S}_{K;N}(\mathbf{z})}p(\mathbf{z} \mid \mathbf{s})\right)
\log |\mathbb{S}_{K;N}(\mathbf{z})|\nonumber\\
\overset{(d)}{=} &\log {N \choose K} - \sum_{(n_1,\ldots,n_K)} \text{Pr}(\mathbb{Z}_{K;N}(n_1,\ldots,n_K)) \log \prod_{k=1}^{K} n_k \nonumber\\
\overset{(e)}{\geq} &\log {N \choose K} - \max_{(n_1,\ldots,n_K)}\log \prod_{k=1}^{K} n_k \label{ap:lem1}\\
\overset{(f)}{\geq}  &\log {N \choose K} - K \log \left(\frac{N}{K}\right),
\end{align}
where $\text{Pr}(\mathbb{Z}_{K;N}(n_1,\ldots,n_K)) = \sum_{\mathbf{z} \in \mathbb{Z}_{K;N}(n_1,\ldots,n_K)} p_{z}(\mathbf{z})$ and $p_z(\mathbf{z})$ is the marginal distribution function with $p(\mathbf{z} \mid \mathbf{s}) \in \mathcal{P}_{z|s}$. In the derivation, line $(a)$ is because of $\mathbf{s} \sim \mathcal{U}(\mathbb{S}_{K;N})$; 
line $(b)$ is because set $\mathbb{Z}_{K;N}$ includes all partition sets $\displaystyle{\mathbb{Z}_{K;N}\left(n_1,\ldots,n_K\right)}$, i.e.,\\ $\displaystyle{\mathbb{Z}_{K;N} = \bigcup_{(n_1,\ldots,n_K)}\mathbb{Z}_{K;N}(n_1,\ldots,n_K)}$;
Line $(c)$ is derived from the log sum inequality, i.e., for non-negative sequence $a_1,\ldots, a_n$ and $b_1,\ldots, b_n$, 
\begin{align}
	\sum_{i=1}^n a_i \log \frac{a_i}{b_i} \geq \left(\sum_{i=1}^n a_i\right) \log \frac{\sum_{j=1}^n a_j}{\sum_{\ell=1}^n b_\ell}
\end{align}
with equality if and only if $\frac{a_i}{b_i}$ is a constant for all $i$. And here, sequence $[a_i] =[p(\mathbf{z}|\mathbf{s})]_{\mathbf{s} \in \mathbb{S}_{K;N}(\mathbf{z})}$, $[b_i] = [\sum_{\tilde{\mathbf{s}} \in \mathbb{S}_{K;N}(\mathbf{z})}p(\mathbf{z} \mid \tilde{\mathbf{s}})]_{\mathbf{s} \in \mathbb{S}_{K;N}(\mathbf{z})}$ is a constant sequence. Line $(d)$ follows for any $\mathbf{z} \in \mathbb{Z}_{K;N}(n_1,\ldots,n_K)$, $|\mathbb{S}_{K;N}(\mathbf{z})| = \prod_{k=1}^{K} n_k$; line (f) is just an application of the inequality of arithmetic and geometric means. For the equality of \eqref{ap:lem1}, line $(c)$, $(e)$ should be equalities, which means by $(b)$, 
\begin{align}
	\frac{p(\mathbf{z} | \mathbf{s})}{\sum_{\tilde{\mathbf{s}} \in \mathbb{S}_{K;N}(\mathbf{z})}p(\mathbf{z} \mid \tilde{\mathbf{s}})} = p(\mathbf{s}| \mathbf{z})=\text{const.}, \quad \forall \mathbf{s} \in \mathbb{S}_{K;N}(\mathbf{z}),
\end{align}
and by $(e)$, 
\begin{align}
	\text{Pr}(\mathbb{Z}_{K;N}(n^*_1,\ldots,n^*_K))=
\begin{cases}
\frac{1}{A}, &(n^*_1,\ldots,n^*_K) = \arg\max \prod_{k=1}^{K} n_k\\
0, &otherwise
\end{cases}
\end{align}
where $A=\sum \text{Pr}(\mathbb{Z}_{K;N}(n^*_1,\ldots,n^*_K))$ is a normalized factor. We can choose $\mathbf{z}|\mathbf{s} \sim \mathcal{U}(\mathbb{Z}_{K;N}(n^*_1,\ldots,n^*_K)\protect\newline \bigcap \mathbb{Z}_{K;N}(\mathbf{s}) )$, and it is easy to see under this condition, both $(b)$ and $(d)$ will be equality, then so is $(e)$. Thus the lower bound  \eqref{ap:lem1} is proved to be achieved. 

	It is worth noting the result won't change with a generalized $\mathbf{z}$. Define $\mathbf{z} \in \{0,1,\ldots,N\}^N$, such that $z_i = 0$ indicates an inactive $i$-th user, and $z_i=k$ indicates that the $i$-th user is assigned into the $k$-th group. The definition of distortion can be generalized as follows:
\begin{align}
	d(\mathbf{s},\mathbf{z}) = 
	\begin{cases}
		0, &\forall i, j \in \mathcal{G}_{\mathbf{s}}, i\neq j, ~\text{and}~z_i, z_j \neq 0, ~\text{and}~z_i s_i \neq z_j s_j\\
		1, &\rm{otherwise}
	\end{cases}.
\end{align}
i.e., active users are assigned different groups, and they cannot be announced as inactive. The definition is consistent with our earlier definition where $\mathbf{z}$ is restricted to $\mathbb{Z}_{K;N}$. From the proof above, it is easy to see with this generalization, the lower bound is the same as in Eq. \eqref{ap:lem1}. The equality in line $(e)$ can be achieved by choosing $\mathbf{z}|\mathbf{s}$ uniformly in the same way.
\end{IEEEproof}
\section{Proof of Theorem \ref{thm1}}\label{appendix.b}
	First, for $\sum {n_k}=N$ and $n_k\geq 0$, define
\begin{align}
	\displaystyle{{N \choose n_1^*,\ldots,n^*_K} = \frac{N!}{\Pi_{k} n_k!}}
\end{align}
to be the number of possible partitions in $\mathbb{Z}_{K;N}(n^*_1,\ldots,n^*_K)$.
\begin{IEEEproof}
	The proof is based on random coding, i.e., randomly generate the source codebook $\mathcal{C} = \{\mathbf{z}_{\ell}\}_{\ell=1}^{L^{(N)}}$ from $p_z(\mathbf{z})$, which is the marginal distribution function based on $p(\mathbf{z} \mid \mathbf{s})$ in \eqref{conditiondis} in Lemma \ref{lem1} and $p_s(\mathbf{s}) =1 /{N \choose K}$, i.e.,
\begin{align}
	p(\mathbf{z} \mid \mathbf{s}) = \frac{1}{K!{N-K \choose n_1^*-1,\ldots,n^*_K-1}},~ \mathbf{z} \in \mathbb{Z}_{K;N}(n^*_1,\ldots,n^*_K)  \bigcap  \mathbb{Z}_{K;N}(\mathbf{s})\\
	p(\mathbf{z}) = \frac{1}{{N \choose n_1^*,\ldots,n^*_K}} , ~ \mathbf{z} \in \mathbb{Z}_{K;N}(n^*_1,\ldots,n^*_K)
\end{align}
	Reveal this codebook to the source encoder and decoder. For any $\mathbf{s} \in \mathbb{S}_{K;N}$, define the source encoding function $f_N^s(\mathbf{s}) = \ell$, such that $\displaystyle{\ell = \arg\min_{1\leq \ell\leq L^{(N)}}d(\mathbf{s},\mathbf{z}_{\ell})}$. If there is more than one such $\ell$, choose the least. Then define the source decoding function $g_N^s(\ell) = \mathbf{z}_{\ell}$. So for any source $\mathbf{s}$, it will be correctly reconstructed if and only if there exists $\ell$ such that $d(\mathbf{s}, \mathbf{z}_{\ell})=0$. The average error probability over the codebook $\mathcal{C}$ is
\begin{align}
	\tilde{P}^{s, (N)}_e = &\sum_{\mathcal{C}} p(\mathcal{C}) \sum_{\mathbf{s}} p(\mathbf{s})\text{Pr}(\forall \mathbf{z}_{\ell} \in \mathcal{C}, d(\mathbf{s},\mathbf{z}_{\ell}) \neq 0 | \mathcal{C},\mathbf{s})\nonumber\\
=&\sum_{\mathcal{C}} p(\mathcal{C}) \sum_{\mathbf{s}:\forall \mathbf{z}_{\ell} \in \mathcal{C}, d(\mathbf{s},\mathbf{z}_{\ell}) \neq 0} p(\mathbf{s})\nonumber\\
=&\sum_{\mathbf{s}}p(\mathbf{s})\sum_{\mathcal{C}:\forall \mathbf{z}_{\ell} \in \mathcal{C}, d(\mathbf{z}^\ell,\mathbf{s})\neq 0} p(\mathcal{C}) \nonumber\\
\overset{(a)}{=}&\sum_{\mathcal{C}:\forall \mathbf{z}_{\ell} \in \mathcal{C}, d(\mathbf{z}^\ell,\tilde{\mathbf{s}})\neq 0} p(\mathcal{C}) \nonumber\\
=&\prod_{\ell=1}^{L^{(N)}} \sum_{\mathbf{z}_{\ell}:d(\mathbf{z}_{\ell},\tilde{\mathbf{s}}) \neq 0} p(\mathbf{z}_{\ell}) \nonumber\\
= &\prod_{\ell=1}^{L^{(N)}} \left(1-\sum_{\mathbf{z}_{\ell}:d(\mathbf{z}_{\ell},\tilde{\mathbf{s}}) = 0} p(\mathbf{z}_{\ell})\right) \nonumber\\
\overset{(b)}{=} &\left(1-K!{N-K \choose n_1^*-1,\ldots,n^*_K-1}\times \frac{1}{{N \choose n_1^*,\ldots,n^*_K}}\right)^{{L^{(N)}}}\nonumber\\
= &\left(1- 2^{-W^I_N}\right)^{L^{(N)}}
\end{align}
The meaning of line $(a)$ is the probability that there are no correct codewords in a random chosen $\mathcal{C}$ for any given $\tilde{\mathbf{s}}$, it is derived by the symmetry of random codebook. Line $(b)$ is derived using the fact that $|\{\mathbf{z}_{\ell}:d(\mathbf{z}_{\ell},\tilde{\mathbf{s}}) = 0\}| = |\mathbb{Z}_{K;N}(n^*_1,\ldots,n^*_K)  \bigcap  \mathbb{Z}(\tilde{\mathbf{s}})|=K!{N-K \choose n_1^*-1,\ldots,n^*_K-1}$. According to the inequality:
\begin{align}
	(1-xy)^n \leq 1-x+e^{-yn}, ~\text{for} ~0\leq x, y\leq 1, n>0
\end{align}
	We have
\begin{align}
	\tilde{P}^{s, (N)}_e \leq e^{-2^{\left(\log L^{(N)}-W_{N}^{I}\right)}}
\end{align}
Since this is the average error probability over all possible codebooks $\mathcal{C}$, there must exist a codebook to achieve the error bound above, which completes the proof.
\end{IEEEproof}
\section{Proof of Lemma \ref{thm3} and Theorem \ref{prop2}}\label{appendix.c}
\begin{proof}
	The proofs of Lemma \ref{thm3} and Theorem \ref{prop2} are similar, so we put them together. In the proof, we will use the method of strong typical set, definition of which can be found in the book of Csiszar and K{\"o}rner \cite{csiszar2011information}. Recall that $T\times 1$ vectors $\mathbf{x}_1$, $\mathbf{x}_2$ represent the codewords of user 1 and 2, let $[\mathbf{x}_1, \mathbf{x}_2]$ denote the $T \times 2$ matrix with $\mathbf{x}_1$ and $\mathbf{x}_2$ as its two columns. A strong typical set $\mathcal{A}_{\epsilon}^{(T)}$ is proposed at first, which contains almost all $[\mathbf{x}_1, \mathbf{x}_2]$, i.e., $\text{Pr}\left([\mathbf{x}_1, \mathbf{x}_2]\in \mathcal{A}_{\epsilon}^{(T)}\right) \xrightarrow{T \to \infty} 1$. In this typical set, the probability of existence of any possible odd cycle (1-odd cycle) is calculated and $P_e$ is bounded by using union bound. 

	To simplify notation, denote $E^{(w)}$, $w=1, 2$ as the event that $H\rq{}$ contains the 1-odd cycles or odd cycles respectively, and $P_e^{(1)} \triangleq P_e^{1-odd} \triangleq \text{Pr}(E^{(1)})$, $P_e^{(2)} \triangleq P_e^{odd} \triangleq \text{Pr}(E^{(2)})$. Since we have:
\begin{align}
	P^{(w)}_e = &\text{Pr}(E^{(w)} [\mathbf{x}_1, \mathbf{x}_2] \in \mathcal{A}_{\epsilon}^{(T)}) + \text{Pr}(E^{(w)}, [\mathbf{x}_1, \mathbf{x}_2] \notin \mathcal{A}_{\epsilon}^{(T)}) \nonumber\\
\leq &\text{Pr}(E^{(w)}, [\mathbf{x}_1, \mathbf{x}_2] \in \mathcal{A}_{\epsilon}^{(T)}) + \text{Pr}([\mathbf{x}_1, \mathbf{x}_2] \notin \mathcal{A}_{\epsilon}^{(T)}),
\label{typicalineq}
\end{align}
it suffices to show that for any $0<p<1$, when $\frac{\log N}{T} <C(p)$, both $\text{Pr}([\mathbf{x}_1, \mathbf{x}_2] \notin \mathcal{A}_{\epsilon}^{(T)})$ and $\text{Pr}(E^{(w)}, [\mathbf{x}_1, \mathbf{x}_2] \in \mathcal{A}_{\epsilon}^{(T)}) $ approach 0 as $N \to \infty$. While the first one is directly from the feature of strong typical set, the key point is to estimate the probability of $E^{(w)}$ in the typical set $\mathcal{A}_{\epsilon}^{(T)}$. First, let us define the strong typical set that will simplify the calculation of $\text{Pr}(E^{(w)}, [\mathbf{x}_1, \mathbf{x}_2] \in \mathcal{A}_{\epsilon}^{(T)})$.

\subsection{Strong typical set}
	Note that the input is $\mathcal{G}_{\mathbf{s}_0} = \{1,2\}$, since the codewords $x_{1,t}$ and $x_{2,t}$ are generated from $\mathcal{B}(p)$, it is very likely that in the set $\left\{(x_{1,t}, x_{2,t})\right\}_{t=1}^{T}$, there are $p_{12}(u,v) T$ pairs $(u,v)$, $\forall u, v\in \{0,1\}$, where $p_{12}(u,v)\triangleq p_x(u)p_x(v)$, and $p_x(\tilde{x})\triangleq p \mathbbm{1}(\tilde{x}=1)+(1-p)\mathbbm{1}(\tilde{x}=0)$ is the pdf of $\mathcal{B}(p)$. Strictly speaking, define $N((u,v)|[\mathbf{x}_1, \mathbf{x}_2])$ as the number of $(u,v)$ in $\left\{(x_{1,t}, x_{2,t})\right\}_{t=1}^{T}$, for any given $\epsilon > 0$, define the strong typical set:
\begin{align} \label{Ae}
	\mathcal{A}_{\epsilon}^{(T)} =\left\{[\tilde{\mathbf{x}}_1, \tilde{\mathbf{x}}_2] \in \{0,1\}^{2T}:\left|\frac{1}{T}N((u,v)|[\tilde{\mathbf{x}}_1, \tilde{\mathbf{x}}_2])-p_{12}(u,v)\right|<\frac{\epsilon}{4},\forall u,v \in \{0,1\}\right\}
\end{align}  
The parameter $\epsilon$ will be chosen at beginning to guarantee some good features of the set $\mathcal{A}_{\epsilon}^{(T)}$, we will provide such requirements on $\epsilon$ during the proof, and summarize them at the end of the proof. The first requirement is that 
\begin{align}
	\epsilon/4 < \max_{u,v} \left(\max(p_{12}(u,v), 1- p_{12}(u,v))\right),\label{e1condition1}
\end{align}
so that $1>p_{12}(u,v)\pm \epsilon/4>0$. For strong typical set, $\forall \epsilon >0$, $\text{Pr}(\mathbf{X} \notin \mathcal{A}_{\epsilon}^{(T)}) \to 0$, as $T\to \infty$, thus let us consider the case that $\text{Pr}(E^{(w)}, [\mathbf{x}_1, \mathbf{x}_2] \in \mathcal{A}_{\epsilon}^{(T)}) \to 0$ in the following parts.

\subsection{Odd cycles for given $\mathcal{A}_{\epsilon}^{(T)}$}	
	Consider $\text{Pr}(E^{(w)}, [\mathbf{x}_1, \mathbf{x}_2] \in \mathcal{A}_{\epsilon}^{(T)})$. For simplicity, assume $N$ is an odd number. Denote $A_M^{(g)}, g \in \{1,2,3\}$ to be the event of existence of the type-$g$ odd cycles with size $M$, thus:
\begin{align}
	E^{(1)} = &\bigcup_{m=1}^{(N-1)/2} A^{(1)}_{2m+1} \\
	E^{(2)} = &\bigcup_{m=1}^{(N-1)/2}  \bigcup_{g=1,2,3} A^{(g)}_{2m+1} 
\end{align}
For any given particular $[\mathbf{x}_1, \mathbf{x}_2] \in \mathcal{A}_{\epsilon}^{(T)}$, denote 
\begin{align}
	P_{2m+1|[\mathbf{x}_1, \mathbf{x}_2]}^{(g)} \triangleq &\text{Pr}\left(A^{(g)}_{2m+1} \big| [\mathbf{x}_1, \mathbf{x}_2]\right),
\end{align}
as the probability that the $g$-th kind of odd cycle of length $2m+1$ exists in $H\rq{}$ for given $[\mathbf{x}_1, \mathbf{x}_2]$. Then by union bound, we have:
\begin{align}
	\text{Pr}(E^{(1)}, [\mathbf{x}_1, \mathbf{x}_2] \in \mathcal{A}_{\epsilon}^{(T)})
\leq &\sum_{[\mathbf{x}_1, \mathbf{x}_2]\in\mathcal{A}_{\epsilon}^{(T)}}\sum_{M=3,5, \ldots, N} P_{M|[\mathbf{x}_1, \mathbf{x}_2]}^{(1)}Q_{1,2}(\mathbf{x}_1,\mathbf{x}_2)\nonumber\\
\leq & \sum_{M=3,5, \ldots, N} \max_{[\mathbf{x}_1, \mathbf{x}_2]\in\mathcal{A}_{\epsilon}^{(T)}} P_{M|[\mathbf{x}_1, \mathbf{x}_2]}^{(1)} \label{PrE1}
\end{align}
where $Q_{1,2}(\mathbf{x}_1,\mathbf{x}_2)$ is the probability that the first two codewords take values $\mathbf{x}_1, \mathbf{x}_2$. Similarly,
\begin{align}
	\text{Pr}(E^{(2)}, [\mathbf{x}_1, \mathbf{x}_2] \in \mathcal{A}_{\epsilon}^{(T)})
\leq \sum_{M=3,5, \ldots, N} \sum_{g=1,2,3}\max_{[\mathbf{x}_1, \mathbf{x}_2]\in\mathcal{A}_{\epsilon}^{(T)}} P_{M|[\mathbf{x}_1, \mathbf{x}_2]}^{(g)} \label{PrE2}
\end{align}
Thus, the key point is to calculate $P_{M|[\mathbf{x}_1, \mathbf{x}_2]}^{(g)}$ for any given $[\mathbf{x}_1, \mathbf{x}_2] \in \mathcal{A}_{\epsilon}^{(T)}$ and upper bound it.

\subsection{The probability of existence of 1-odd cycles of length $M$: $P_{M|[\mathbf{x}_1, \mathbf{x}_2]}^{(1)}$}
	Consider any particular 1-odd cycle of length $M$, denoted by $H^{(1)}_{e;M}=(1,2,i_1,\ldots, i_{M-2})$, there are at most ${N-2 \choose M-2}(M-2)!$ such odd cycles out of $N$ nodes, and because of symmetry, the existence of any of them is equiprobable. Let us see for a given $[\mathbf{x}_1, \mathbf{x}_2]$, what values should the codewords $\{\mathbf{x}_{i_1}, \ldots, \mathbf{x}_{i_{M-2}}\}$ of other $M-2$ vertices be to guarantee $H^{(1)}_{e;M} \subseteq H\rq{}$. 

	Note $H\rq{}$ is obtained by a series of operations on graph, and the order of operations is not important. Let us calculate the probability that $H^{(1)}_{e;M}$ is not deleted at any slot $1\leq t \leq T$. By the symmetry of the generation of codewords, this probability only depends on the values of $(x_{1,t}, x_{2,t})$. Given $[\mathbf{x}_1, \mathbf{x}_2]$, let $\mathcal{T}_{u,v} = \{t:(x_{1,t},x_{2,t})=(u,v)\}$, and $T_{u,v} \triangleq |\mathcal{T}_{u,v}| = N((u,v)|[\mathbf{x}_1, \mathbf{x}_2])$. Thus we just need to consider four situations $t \in \mathcal{T}_{u,v}$. Denote the probability that $H^{(1)}_{e;M}$ is not deleted at $t \in \mathcal{T}_{u,v}$ as $\mu^{(1)}_{u,v;M}$, so by union bound of all possible $H^{(1)}_{e;M}$, we have for all $[\mathbf{x}_1, \mathbf{x}_2] \in \mathcal{A}_{\epsilon}^{(T)}$:
\begin{align}
	P_{M|[\mathbf{x}_1, \mathbf{x}_2]}^{(1)} \leq  &\displaystyle{{N-2 \choose M-2}(M-2)! \Pi_{u, v} \left(\mu^{(1)}_{u,v;M}\right)^{T_{u,v}}} \nonumber\\
\leq &\displaystyle{N^{M-2} \Pi_{u, v} \left(\mu^{(1)}_{u,v;M}\right)^{(p_{12}(u,v)-\epsilon/4)T}} \label{boundpm1}
\end{align}
	Now $\mu^{(1)}_{u,v;M}$ is determined separately for cases of different $(u,v)$ as follows:

\begin{enumerate}
	\item The case that $t \in \mathcal{T}_{0,0}$

	For $t \in \mathcal{T}_{0,0}$, $y_t=0$, the operation is to delete vertices. Then all of the codewords of other $M-2$ vertices should be 0, otherwise these vertices would be deleted, thus
\begin{align}
	\mu^{(1)}_{0,0;M} = \text{Pr}(x_{i_w, t} = 0, \forall w \in \{1,\ldots, M-2\})=(1-p)^{M-2}
\end{align}

\item The case that $t \in \mathcal{T}_{1,1}$

	In these slots $y_t=1$, the operation of deleting clique is done in $H\rq{}$. At any slot $t$, $H^{(1)}_{e;M}$ will be broken up if the edges of it are deleted, which is equivalent to the existence of codewords of two consecutive vertices from $\{1, 2, i_1, \ldots, i_{M-2},1\}$ to be both 0 at those $t\in \mathcal{T}_{1,1}$, as shown in Fig. \ref{pfig14}. So we have:


\begin{figure}
\centering
\includegraphics[width=2in]{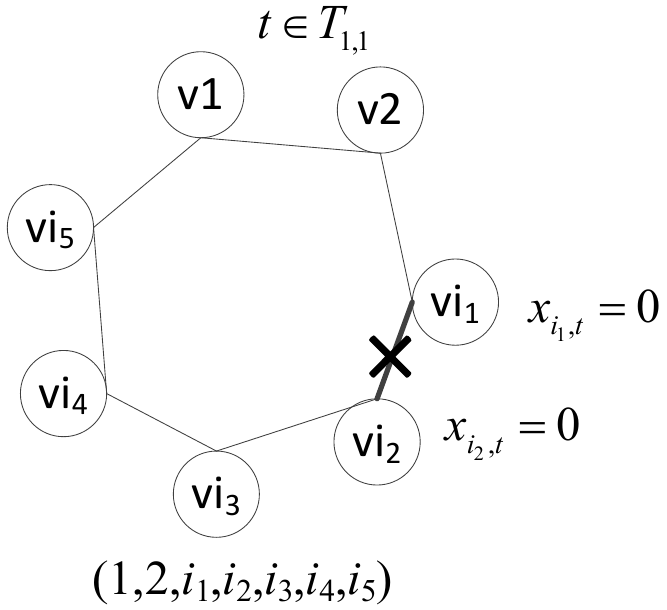}
\caption{Example that odd cycle is destroyed by edge $(i_1,i_2)$ deleted.} \label{pfig14}
\end{figure}
\begin{figure}
\centering
\includegraphics[width=1.5in]{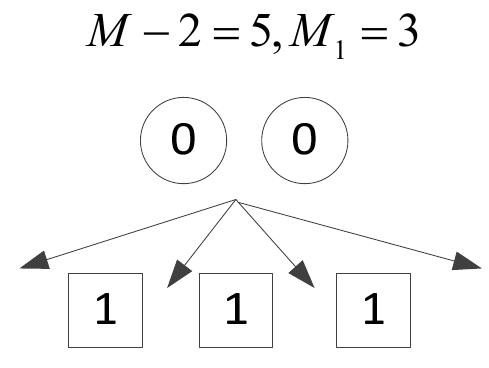}
\caption{Example of counting number of sequence $\{{x}_{i_1, t}, \ldots, x_{i_{M-2}, t}\}$ with $M_1$ ones, here $M-2=5$, $M_1=3$.} \label{pfig15}
\end{figure}
\begin{align}
	\mu^{(1)}_{1,1;M} \triangleq & 1 - \text{Pr}\left(\exists (i,j) \in \{(1,2), (2, i_1), \ldots, (i_{M-3}, i_{M-2}), (i_{M-2},1)\}, x_{i, t} = x_{j, t} =0\right)\nonumber\\
\overset{(a)}{=} & 1 - \text{Pr}\left(\exists w \in \{1,\ldots, M-3\}, {x}_{i_w, t} = {x}_{i_{w+1}, t} =0\right)\nonumber\\
\overset{(b)}{=}&1 - \sum_{M_1=\frac{M-3}{2}}^{M-2}\text{Pr}\left(\sum_{w=1}^{M-2}{{x}_{i_w, t}} = M_1, \exists w \in \{1,\ldots, M-3\}, {x}_{i_w, t} = {x}_{i_{w+1}, t} =0\right)\nonumber\\
\overset{(c)}{=}&\sum_{M_1=\frac{M-3}{2}}^{M-2} {M_1+1 \choose M-2-M_1} p^{M_1} (1-p)^{M-2-M_1}.
\label{x1}
\end{align}
Line $(a)$ is because now $x_{1,t}=x_{2,t}=1$. For line $(b)$, we change the sum by grouping items with different $M_1$, where $M_1$ is the number of values of $1\leq w \leq M-2$ for which $x_{i_w,t}=1$. It is easy to see there must be $M_1 \geq \lfloor\frac{M-2}{2}\rfloor=\frac{M-2}{2}$, otherwise there must exists a $w$ such that $x_{i_{w}, t} = x_{i_{w+1}, t} =0$. In line $(c)$, the sum of probability of the items with $M_1$ is calculated out, since the probability of each item is $p^{M_1} (1-p)^{M-2-M_1}$, the key point is to count the number of sequences $({x}_{i_1, t}, \ldots, x_{i_{M-2}, t})$ with $M_1$ ones and $M-2-M_1$ zeros. The counting method is to fix $M_1$ ones, then count how many combinations to put $M-2-M_1$ zeros into $M_1+1$ spots, as shown in Fig. \ref{pfig15}. 

	For a better statement in the rest of the paper, define a function $J_M(p)$ which calculates the probability of a $M$ length random Bernoulli sequence $(x_1, \ldots, x_M)$, $x_w \sim \mathcal{B}(p)$ without consecutive zeros, i.e., 
\begin{align}
	J_M(p) = &1 - \text{Pr}\left(\exists w \in \{1,\ldots, M-1\}, x_w = x_{w+1} =0\right)\nonumber\\
=&\sum_{M_1=\lfloor\frac{M}{2}\rfloor}^{M} {M_1+1 \choose M-M_1} p^{M_1} (1-p)^{M-M_1}.
\label{x1}
\end{align}
Thus, we can see $\mu^{(1)}_{1,1;M} = J_{M-2}(p)$.
\item The cases that  $t\in \mathcal{T}_{1,0}$ or $t\in \mathcal{T}_{0,1}$
	
	The two cases are symmetric, let's consider $t\in \mathcal{T}_{1,0}$ first. It is similar as the case that $t\in \mathcal{T}_{1,1}$, but since now ${x}_{2,t} = 0$, the codeword of vertex that connected to vertex $2$ (i.e., vertex $i_1$), should be ${x}_{i_1,t} = 1$. For other $M-3$ vertices, it is required codewords of any two consecutive vertices from $\{ i_2, \ldots, i_{M-2}\}$ at those $t \in \mathcal{T}_{1,0}$ should not be both 0. Thus,
\begin{align}
	\mu^{(1)}_{1,0;M} = &\text{Pr}({x}_{i_1 t} = 1)\left(1 - \text{Pr}\left(\exists w \in \{1,\ldots, M-1\}, x_w = x_{w+1} =0\right)\right)\nonumber\\
=&p J_{M-3}(p)
\end{align}
 Similarly, for $t\in \mathcal{T}_{0,1}$, we have $\mu^{(1)}_{1,0;M} = J_{M-3}(p)$.
\end{enumerate}
	Then $P_{M|[\mathbf{x}_1, \mathbf{x}_2]}^{(1)}$ can be bounded by \eqref{boundpm1}.

\subsection{The probability of existence of type-2 and type-3 cycles of length $M$: $P_{M|[\mathbf{x}_1, \mathbf{x}_2]}^{(2)}$ and $P_{M|[\mathbf{x}_1, \mathbf{x}_2]}^{(3)}$}
	For the type-2 odd cycles, either 1 or 2 nodes are included. Denote $H^{(2),h}$ to be a second kind odd cycle containing vertex $h \in \{1,2\}$. We can choose a particular odd cycle $H^{(2),1}_{e;M}=(1,i_1,\ldots, i_{M-1})$, $H^{(2),2}_{e;M}=(2,i_1,\ldots, i_{M-1})$, and a particular type-3 odd cycle $H^{(3)}_{e;M}=(i_1,\ldots, i_{M})$. There are ${N-2 \choose M-1}\frac{(M-1)!}{2}$ such $H^{(2), h}_{e;M}$, and ${N-2 \choose M}\frac{(M-1)!}{2}$ such $H^{(3)}_{e;M}$. Then following the same analysis as for $P_{M|[\mathbf{x}_1, \mathbf{x}_2]}^{(1)}$, we have for all $[\mathbf{x}_1, \mathbf{x}_2] \in \mathcal{A}_{\epsilon}^{(T)}$,
\begin{align}
	P_{M|[\mathbf{x}_1, \mathbf{x}_2]}^{(2)} \leq &\sum_{h=1,2} {N-2 \choose M-1}\frac{(M-1)!}{2} \Pi_{u, v} \left(\mu^{(2), h}_{u,v;M}\right)^{T_{u,v}}\nonumber\\
\leq &\frac{1}{2}N^{M-1} \sum_{h=1,2}\Pi_{u, v} \left(\mu^{(2), h}_{u,v;M}\right)^{(p_{12}(u,v)-\epsilon/4)T},
 \label{boundpm2}
\end{align}
\begin{align}
	P_{M|[\mathbf{x}_1, \mathbf{x}_2]}^{(3)} \leq &{N-2 \choose M}\frac{(M-1)!}{2} \Pi_{u, v} \left(\mu^{(3)}_{u,v;M}\right)^{T_{u,v}}\nonumber\\
\leq &N^{M} \Pi_{u, v} \left(\mu^{(3)}_{u,v;M}\right)^{(p_{12}(u,v)-\epsilon/4)T}, \label{boundpm3}
\end{align}
where $\mu^{(g),h}_{u,v;M}$ is the probability that $H^{(g),h}_{e;M}$ won\rq{}t be deleted at $t \in \mathcal{T}_{u,v}$. Then similarly, we have:
\begin{enumerate}
	\item For $t \in \mathcal{T}_{0,0}$, $y_t=0$, every vertex cannot be deleted, thus:
\begin{align}
	\mu^{(2),h}_{0,0;M} = (1-p)^{M-1};\quad \mu^{(3)}_{0,0;M} = (1-p)^{M}
\end{align}
	\item For $g=2$, let\rq{}s consider $H^{(2),1}_{e;M}$ first, when $t \in \mathcal{T}_{1,1}$ or $t \in \mathcal{T}_{1,0}$, $x_{1,t}=1$, thus:
\begin{align}
	\mu^{(2),1}_{1,0;M}=\mu^{(2),1}_{1,1;M} = 1 - \text{Pr}\left(\exists w \in \{1,\ldots, M-2\}, x_{i_{w}, t} = x_{i_{w+1}, t} =0\right)=J_{M-1}(p)
\end{align}
When $t \in \mathcal{T}_{0,1}$, $x_{1,t}=0$, thus, the codewords of vertexs $i_1$ and $i_{M-1}$ which are connected to $1$ should be 1, i.e., 
\begin{align}
	\mu^{(2),1}_{0,1;M}= &\text{Pr}(x_{i_1,t}=x_{i_{M-1},t}=1)\left(1 - \text{Pr}\left(\exists w \in \{2,\ldots, M-1\}, x_{i_{w}, t} = x_{i_{w+1}, t} =0\right)\right)\nonumber\\
=&p^2 J_{M-3}(p)
\end{align}
	For $H^{(2),2}_{e;M}$, due to symmetry, the result is easy to derive:
\begin{align}
	\mu^{(2),2}_{1,1;M}=\mu^{(2),2}_{0,1;M}=J_{M-1}(p); \quad
	\mu^{(2),2}_{1,0;M}=p^2 J_{M-3}(p)
\end{align}
\item For $g=3$, for $t \notin \mathcal{T}_{0,0}$, $y_t=1$. Since neither vertices 1 nor 2 are in $H^{(3)}_{e;M}$, $\mu^{(3)}_{u,v;M}$ are the same for any $(u,v)\neq (0,0)$. Now we have:
\begin{align}
	\mu^{(3)}_{u,v;M} = &1 - \text{Pr}\left(\exists (i,j) \in \{(i_1, i_2), \ldots, (i_{M-1}, i_{M}), (i_{M},i_1)\}, x_{i, t} = x_{j, t} =0\right)\nonumber\\
=&1 - \text{Pr}\left(\exists w \in \{1, \ldots, M-1\}, x_{i_w, t} = x_{i_{w+1}, t} =0\right) \nonumber\\
&
-\text{Pr}\left((x_{i_1, t},x_{i_M, t})=(0,0); \nexists w \in \{1, \ldots, M-1\}, x_{i_w, t} = x_{i_{w+1}, t} =0\right)\nonumber\\
= &J_M(p) - \text{Pr}((x_{i_1, t},x_{i_M, t})=(0,0), x_{i_2, t}=1, x_{i_{M-1}, t}=1)\times\nonumber\\
 &\left(1- \text{Pr}\left(w \in \{3, \ldots, M-3\}, x_{i_w, t} = x_{i_{w+1}, t} =0\right)\right)\nonumber\\
=&J_M(p) - p^2(1-p)^2J_{M-4}(p)
\end{align}
\end{enumerate}
	
	Now we can bound $P_{M|[\mathbf{x}_1, \mathbf{x}_2]}^{(g)}$. In the next subsection, we will get explicit expressions for $J_M(p)$, then $\mu^{(g)}_{u,v;M}$. We can see they have a close relationship to extended Fibonacci numbers.

\subsection{Explicit expressions for $J_M(p)$ and extended Fibonacci numbers}
	We will show that expression for $J_M(p)$ has a close relation to a certain extended Fibonacci numbers. It is not surprising since Fibonacci numbers can be used for determining the numbers of consecutive 0s in Bernoulli sequence \cite{koshy2011fibonacci}.

Define extended Fibonacci numbers as:
\begin{align}
	F(k,p) = \sum_{j=0}^{\lfloor \frac{k-1}{2} \rfloor} F(k,j) p^{k-1-j}(1-p)^j\\
	F(k,j) = 
\begin{cases}
{k-1-j \choose j}, &\quad 0\leq j \leq  \lfloor \frac{k-1}{2} \rfloor \\
0,	&otherwise
\end{cases}
\end{align}
The meaning of $F(k,j)$ can be seen directly from the Pascal triangle, as shown in Fig. \ref{pfig10}, and $F(k,p)$ is a weighted sum of $F(k,j)$ with the weight $p^{k-1-j}(1-p)^j$. From Fig. \ref{pfig10} it is shown:


\begin{figure}
\centering
\includegraphics[width=1.8in]{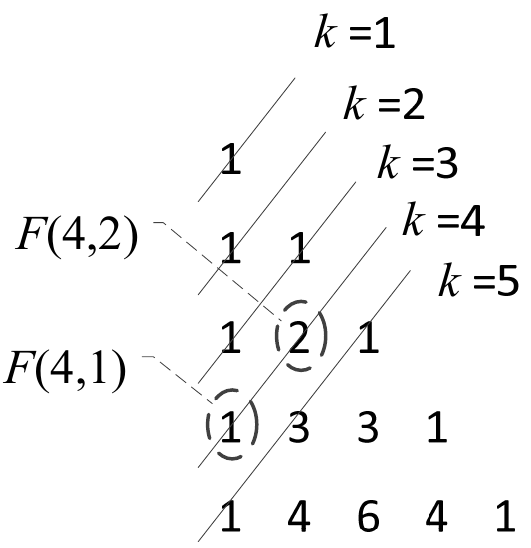}
\caption{$F(k,j)$ in the Pascal triangle.} \label{pfig10}
\end{figure}

\begin{align}
	F(k,j) = F(k-1,j) + F(k-1,j-1),
\end{align}
so that
\begin{align}
	F(k,p) &= \sum_{j=0}^{\lfloor \frac{k-1}{2} \rfloor} \left(F(k-1,j) + F(k-1,j-1)\right) p^{k-1-j}(1-p)^j \nonumber\\
&= pF(k-1,p) + p(1-p)F(k-2,p).
\end{align}
Then we can get the general terms of $F(k,p)$ by solving the corresponding difference equation, which gives us:
\begin{align}
	F(k,p) = \frac{\varphi(p)^k - \psi(p)^k}{\varphi(p) - \psi(p)},
\label{fkp}
\end{align}
where 
\begin{align}
	\varphi(p) = \frac{p+\sqrt{4p-3p^2}}{2}; \quad \psi(p) = \frac{p-\sqrt{4p-3p^2}}{2}
\end{align}
	It is not difficult to see
\begin{align}
	1\geq \varphi(p)\geq 0 \geq \psi(p) \geq-1,~|\varphi(p)|\geq |\psi(p)|
\end{align}
	Given $F(k,p)$ defined in \eqref{fkp}, it is straightforward to see:
\begin{align}
	J(M,p) = \frac{1}{p} F(M+2,p) =\frac{1}{p} \frac{\varphi(p)^{M+2} - \psi(p)^{M+2}}{\varphi(p) - \psi(p)}
\end{align}
   Which further enables us to determine $\mu^{(g)}_{u,v;M}$ and  upperbound $P_{M|[\mathbf{x}_1, \mathbf{x}_2]}^{(g)}$.

\subsection{Bounds of $P_{M|[\mathbf{x}_1, \mathbf{x}_2]}^{(g)}$}
	By Eq. \eqref{boundpm1}, \eqref{boundpm2}, \eqref{boundpm3}, now we have for any $[\mathbf{x}_1, \mathbf{x}_2] \in \mathcal{A}_{\epsilon}^{(T)}$, 
\begin{align}
	P_{M|[\mathbf{x}_1, \mathbf{x}_2]}^{(g)} \leq 2^{-(M-3+g)T\left(\left(h^{(g)}- ((1-p)^2-\frac{\epsilon}{4}) \log (1-p)\right) - \frac{\log N}{T}\right)} \label{boundpe}
\end{align}
where
\begin{align}
	h^{(1)} \triangleq &-\frac{1}{M-2}\left((p^2-\frac{\epsilon}{4})\log J_{M-2}(p) + (2p(1-p) - \frac{\epsilon}{2}) \log pJ_{M-3}(p) \right)\\
	h^{(2)} \triangleq &-\frac{1}{M-1}\left((p-\frac{\epsilon}{2})\log J_{M-1}(p) + (p(1-p) - \frac{\epsilon}{4}) \log p^2J_{M-3}(p) \right)\\
	h^{(3)} \triangleq &-\frac{1}{M}\left(1-(1-p)^2-\frac{3\epsilon}{4}\right)\log \left(J_{M}(p)-p^2(1-p)^2J_{M-4}(p)\right)
\end{align}
	Next we will give a concise lower bound of $h^{(g)}$, which can be obtained by the monotonicity and concavity of $\log(\cdot)$.

(1) Bound of $h^{(1)}$

	Define a normalizing factor
\begin{align}
	W = (p^2-\frac{\epsilon}{4}) + (2p(1-p) - \frac{\epsilon}{2}) = 1-(1-p)^2-\frac{3\epsilon}{4}.
\end{align}
	If we choose $\epsilon$ so that 
\begin{align}
	W - \varphi(p)^2 = \frac{p}{2} \left((2-p) - \sqrt{(2-p)^2-4(1-p)^2}\right) -\frac{3\epsilon}{4} >0. \label{e1condition2}
\end{align}
then when $M$ is an odd number,
\begin{align}
-h^{(1)}\overset{(a)}{\leq} & W\log \left(\frac{p^2-\epsilon/4}{W}J_{M-2}(p) + \frac{2p(1-p) -\epsilon/2}{W}pJ_{M-3}(p)\right) \nonumber\\
\overset{(b)}{\leq} &W\log \left(\frac{p^2J_{M-2}(p)+ 2p^2(1-p)J_{M-3}(p)}{W}\right) \label{1-odd1}\\
=&W\log \left(\frac{p\sqrt{4-3p}}{W(\varphi(p)-\psi(p))}\left(\frac{\sqrt{4-3p}+\sqrt{p}}{2} \varphi(p)^{M-1}- \frac{\sqrt{4-3p}-\sqrt{p}}{2} \psi(p)^{M-1} \right) \right)\nonumber\\
\overset{(c)}{\leq} &W\log \left(\frac{p}{W(\varphi(p)-\psi(p))}\frac{\sqrt{4-3p}(\sqrt{4-3p}+\sqrt{p})}{2} \varphi(p)^{M-1}\right) \nonumber\\
=&W \log\left(\frac{\varphi(p)^{M}}{W}\right)\label{eq:h12}\\
=&W \log\left(\varphi(p)^{M-2}\right) + W \log\left(\frac{\varphi(p)^{2}}{W}\right)\nonumber\\
\leq &W \log\left(\varphi(p)^{M-2}\right)
\end{align}
Where line $(a)$ is because of the concavity of $\log(\cdot)$; inequalities $(b)$ and $(c)$ are because of the increasing monotonicity of $\log(\cdot)$. Then 
\begin{align}
	h^{(1)} -((1-p)^2-\frac{\epsilon}{4}) \log (1-p) \geq C(p) - g_1(p) \epsilon
\end{align}
where $g_1(p) \triangleq - \frac{1}{4}\left(3\log (\varphi(p)) +\log (1-p) \right) >0$, and 
\begin{align}
	C(p) \triangleq -(1-(1-p)^2)\log \varphi(p) -(1-p)^2\log (1-p).
\label{cp_app}
\end{align}

	(2) Bound of $h^{(2)}$

	Similarly, if $\epsilon$ satisfies \eqref{e1condition1} and \eqref{e1condition2}, we have 
\begin{align}
	-h_2 
\overset{(a)}{\leq} &W \log \left(\frac{p - \epsilon/2}{W}J_{M-1}(p) + \frac{p(1-p)-\epsilon/4}{W} p J_{M-3}(p)\right) + (p(1-p)-\epsilon/4) \log p \nonumber\\
\overset{(b)}{\leq} &W \log \left(\frac{p}{W}J_{M-1}(p) + \frac{p(1-p)}{W} p J_{M-3}(p)\right) + (p(1-p)-\epsilon/4) \log p\nonumber\\
\overset{(c)}{\leq} &W \log\left(\frac{\varphi(p)^{M}}{W}\right) +(p(1-p)-\epsilon/4) \log p\nonumber\\
= & W \log\left(\varphi(p)^{M-1}\right) + W\log\left(\frac{\varphi(p)}{W}\right) +(p(1-p)-\epsilon/4) \log p\nonumber\\
\overset{(d)}{\leq} & W \log\left(\varphi(p)^{M-1}\right) -\frac{\epsilon}{4} \log p
\end{align}
where inequalities $(a)$ and $(b)$ are because of the concavity and monotonicity of $\log(\cdot)$; line $(c)$ is because that $\displaystyle{\frac{p}{W}J_{M-1}(p) + \frac{p(1-p)}{W} p J_{M-3}(p) = \frac{p^2J_{M-2}(p)+ 2p^2(1-p)J_{M-3}(p)}{W}}$, and then is the same with \eqref{1-odd1} and \eqref{eq:h12}; line $(d)$ is derived from Eq. \eqref{e1condition2}, and: 
\begin{align}
	&W\log\left(\frac{\varphi(p)}{W}\right) +(p(1-p)-\epsilon/4) \log p\nonumber\\
=&\left(\frac{W}{2} \log (\varphi(p)^2/W) -  (2p-p^2) \log \sqrt{2p-p^2} + p(1-p) \log p\right) \nonumber\\
	&+\left(\frac{3\epsilon}{8} \log (1-(1-p)^2) + \frac{W}{2}\log \left(1- \frac{3\epsilon}{4(2p-p^2)}\right) -\frac{\epsilon}{4} \log p\right) \nonumber\\
\leq &\left(-  (2p-p^2) \log \sqrt{2p-p^2} + p(1-p) \log p\right) -\frac{\epsilon}{4} \log p\nonumber\\
=& -p(1-H(p/2)) -\frac{\epsilon}{4} \log p\nonumber\\
\leq &-\frac{\epsilon}{4} \log p
\end{align}
and thus,
\begin{align}
	h_2 - ((1-p)^2-\frac{\epsilon}{4})
\geq &-W \log \varphi(p) - ((1-p)^2-\frac{\epsilon}{4}) \log (1-p) +\frac{\epsilon \log p}{4(M-1)} \nonumber\\
= & C(p) + \left(3\log (\varphi(p)) +\log (1-p) \right) \frac{\epsilon}{4} +\frac{\epsilon \log p}{4(M-1)} \nonumber\\
\geq &C(p) -g_2(p) \epsilon,
\end{align}
where $g_2(p) = g_1(p)-\log p/8>0$.

	(3) Bound of $h^{(3)}$

	Similarly, we can bound $h_3$ if $\epsilon$ satisfies \eqref{e1condition1} and \eqref{e1condition2}, and $M$ is odd number,
\begin{align}
	&h_3 - ((1-p)^2-\frac{\epsilon}{4}) \log (1-p) 
 \nonumber\\
\geq&-\frac{1}{M}\left(1-(1-p)^2-\frac{3\epsilon}{4}\right)\log (\varphi(p)^M + \psi(p)^M) - ((1-p)^2-\frac{\epsilon}{4}) \log (1-p)\nonumber\\
\overset{(a)}{\geq} &\frac{1}{M}\left(1-(1-p)^2-\frac{3\epsilon}{4}\right)\log (\varphi(p)^M) - ((1-p)^2-\frac{\epsilon}{4}) \log (1-p)\nonumber\\
=& C(p) - g_1(p)  \epsilon
\end{align}
where line $(a)$ is because $\psi(p) < 0$ and $M$ is odd number. 

	As determined above, since $g_2(p) > g_1(p)$, we can summarize the results that when $\epsilon$ satisfies \eqref{e1condition1} and \eqref{e1condition2}, we have:
\begin{align}
	h^{(g)} - ((1-p)^2-\frac{\epsilon}{4}) \log (1-p) \geq C(p) - g_2(p) \epsilon
\end{align}
and thus from Eq. \eqref{boundpe}, we have 
\begin{align}
	\displaystyle{\max_{[\mathbf{x}_1, \mathbf{x}_2] \in \mathcal{A}_{\epsilon}^{(T)}}P_{M|[\mathbf{x}_1, \mathbf{x}_2]}^{(g)} \leq 2^{-(M-3+g)T\left(C(p)-g_2(p)\epsilon - \frac{\log N}{T}\right)}}. \label{pembound}
\end{align}

\subsection{Completing the proof}
	If $\frac{\log N}{T} \leq C(p) -\xi$ for any constant $\xi>0$, we can always choose $\epsilon$ satisfying \eqref{e1condition1} and \eqref{e1condition2}, and
\begin{align}
	\epsilon < (C(p) - \delta)/g_2(p) 
\end{align}
so that $C(p) - \frac{\log N}{T} -g_2(p)\epsilon \geq C(p) - \delta -g_2(p)\epsilon \triangleq \Delta >0$, where $\Delta$ is a predetermined constant. Then by Eq. \eqref{PrE1}, \eqref{PrE2} and \eqref{pembound}, we have:
\begin{align}
	\text{Pr}(E^{(1)}, [\mathbf{x}_1, \mathbf{x}_2] \in \mathcal{A}_{\epsilon}^{(T)})
\leq &\sum_{M=3,5, \ldots, N} \max_{[\mathbf{x}_1, \mathbf{x}_2]\in\mathcal{A}_{\epsilon}^{(T)}} P_{M|[\mathbf{x}_1, \mathbf{x}_2]}^{(1)}\nonumber\\
\leq  &\sum_{M=3,5, \ldots, N} 2^{-(M-2) \Delta T}\nonumber\\
\leq &\frac{2^{-\Delta T}}{1-2^{-2\Delta T}} 
\end{align}
and 
\begin{align}
	\text{Pr}(E^{(2)}, [\mathbf{x}_1, \mathbf{x}_2] \in \mathcal{A}_{\epsilon}^{(T)})
\leq &\sum_{g=1,2,3}\sum_{M=3,5, \ldots, N} \max_{[\mathbf{x}_1, \mathbf{x}_2]\in\mathcal{A}_{\epsilon}^{(T)}} P_{M|[\mathbf{x}_1, \mathbf{x}_2]}^{(g)}\nonumber\\
\leq  &\sum_{g=1,2,3}\sum_{M=3,5, \ldots, N} 2^{-(M-3+g) \Delta T}\nonumber\\
\leq &3 \times \frac{2^{-\Delta T}}{1-2^{-2\Delta T}} 
\end{align}
Thus, when $N \to \infty$, which also means $T \to \infty$, we can see $\text{Pr}(E^{(w)}, [\mathbf{x}_1, \mathbf{x}_2] \in \mathcal{A}_{\epsilon}^{(T)})$ approaches 0, $\forall w=1, 2$. Since $\text{Pr}(E^{(w)}, [\mathbf{x}_1, \mathbf{x}_2] \in \mathcal{A}_{\epsilon}^{(T)}) \to 0$ as well, we have $P_e^{(w)} \to 0$ when $\frac{\log N}{T} < C(p) -\xi$, which completes the proof.
\end{proof}
\bibliographystyle{IEEEtran}
\bibliography{toword}

\end{document}